\documentstyle[preprint,eqsecnum,aps]{revtex}
\begin{document}
\draft
\title{Transport processes in anisotropic gravitational collapse}
\author{J. Mart\'{\i}nez}
\address{Physics Department, Faculty of Sciences, Building Cc,\\
Universidad Aut\'onoma de Barcelona,\\ 08193, Bellaterra, Barcelona, Spain.}
\maketitle

\begin{abstract}
In this paper we introduce a new method to study the influence of thermal
conduction and viscous processes in anisotropic gravitational collapse. To
this end we employ the HJR method to solve the Einstein equations. The
Maxwell-Cattaneo type transport equations are used to find the temperature
and bulk and shear viscous pressures. Under some conditions Maxwell-Cattaneo
transport equations comply with relativistic causality. Thus, it is
advisable to use them instead of Eckart transport equations. In the inner
layers of the star the temperature ceases to be sensitive to the boundary
condition. This behavior, which can be explained in terms of the Eddington
approximation, allows us to find the thickness of the neutrinosphere. The
dynamics of collapsing dense stars is deeply influenced by the neutrino
emission/absorption processes. These cool the star and drive it to a new
equilibrium state. Therefore, the calculation of transport coefficients is
based on these processes.
\end{abstract}

\section{INTRODUCTION}

The physical study of relativistic stars rests usually on the assumption of
local isotropy. However, at least for high densities, theoretical evidence
suggests that this may not be a very accurate approximation \cite
{Canuto74,BoLi74}. Then, it seems fitting to adopt anisotropic models to
carry out a more detailed analysis of the internal structure of neutron
stars, their upper-mass limit and their collapse dynamics. Because of the
anisotropy, shear viscous stresses arises. The interchange of momentum
between the different layers of the star causes viscous processes. It
unavoidably comes from the interaction between particles and from the
matter-radiation interaction as well. Therefore, the presence of radiation
flows during the collapse reveals the existence of anisotropy in the system.
When radiation and matter interact strongly (i.e., diffusion) shear and bulk
viscous pressure arise. On the opposite limit (free streaming) one has
anisotropy associated to the direction of propagation of the radiation.
Because of this, it is suitable to introduce an anisotropic state equation
if one wishes to analyze systems in which the flow of radiation non
negligibly influences the dynamics of the collapse \cite
{HeJiBa89,BaRo92,MaPaNu94}. It is then important to study the anisotropy
associated to the radiation flow during the collapse.

We shall adopt the semi numerical method of Herrera et al. \cite{HeJiRu80} -
HJR henceforth - to solve the Einstein equations. Furthemore, to find the
temperature and bulk and shear viscous pressure it will be necessary to
introduce a set of transport equations. To solve them we will adopt the
explicit expressions for the transport coefficients of heat conductivity and
bulk and shear viscosity of an interacting mixture of matter and radiation
found by Weinberg \cite{Weinberg71}. This requires to determine which
reactions are more relevant in the momentum transport between the different
star layers. The large neutrino mean free path implies high transport
coefficients. Thus, viscous processes induced by interactions between
neutrinos and matter are much more important than those induced by other
scattering processes. The cooling by absorption and emission of neutrinos
drive the star to the equilibrium. Among the sources of neutrino opacities
it seems that the scattering with electrons ($e^{-}+\nu \rightarrow
e^{-}+\nu $) and nucleon absorption ($\nu +n\rightarrow e^{-}+p$) may
account for this process \cite{ShTe83,Bruenn85,Demianski85,Bahcall89}.

The temperature is a key quantity to decide which processes can take place
during the collapse. Moreover it is indispensable to calculate the bulk and
shear viscous pressure. Unluckily the solution of the Einstein equations
does not provide any information about it. This is why a set of transport
equations must be adopted. The standard Eckart's theory of irreversible
processes \cite{Eckart40,KlGrMa53,LaLi71} exhibits undesirable effects
(i.e., it predicts an infinite speed of propagation for thermal and viscous
signals and unstable equilibrium states). It is then necessary to resort to
another thermodynamic theory of irreversible processes that does not present
this anomalous behavior. The {\it extended irreversible thermodynamics}
theory (EIT) \cite{Israel76,PaJoCa82} seems to be a good candidate to
replace the old Eckart theory and it has been used in cosmological problems
with good results \cite{PaBaJo91,CaSa92,RoPa93}. Essentially the EIT rests
on two hypotheses: (1) The dissipative flows (heat flow and viscous
pressures) are considered as independent variables, hence, the entropy
function depends not only on the classical variables (particle number and
energy density) but on these dissipative flows as well. (2) At equilibrium
state, the entropy function is a maximum. Moreover, its flow depends on all
dissipative flows and its production rate is semipositive definite. As a
consequence the heat flow ${\cal Q}^\mu $, the bulk viscous pressure $\Pi $
and the traceless viscous tensor $\pi _{\mu \nu }$ obey the evolution
equations \cite{JoCaLe93} 
$$
\tau _{{\cal Q}}\stackrel{\cdot }{{\cal Q}^\nu }h_\nu^\mu+{\cal Q}^\mu 
=\chi h^{\mu\nu }\left[ T_{,\nu }-T\stackrel{\cdot }{U}_\nu -
T\left( \alpha_0 \Pi _{,\nu}-\alpha_1 \Pi _{\nu ;\rho }^\rho 
\right) \right] , 
$$
$$
\tau _\Pi U^\alpha \Pi_{,\alpha}+\Pi =-\zeta U_{;\mu }^\mu +\alpha_0 \zeta 
{\cal Q}_{;\mu }^\mu  
$$
and%
$$
\tau _\pi \stackrel{\cdot }{\pi }_{\mu \nu }+\pi _{\mu \nu }=
2\eta\sigma_{\mu\nu}-2\eta\alpha_1{\cal{Q}}_{\alpha;\beta}
\left(h^\alpha_{(\mu}h^\beta_{\nu)}-\frac{1}3h_{\mu\nu}h^{\alpha\beta}
\right)
$$
respectively, where $\tau _{{\cal Q}}$, $\tau _\Pi $ and $\tau _\pi $ are
the relaxation time of thermal, bulk viscous and shear viscous signals
respectively. $T$ stands for the temperature, $\chi $ for the thermal
conductivity coefficient, and $\zeta $ and $\eta $ for the bulk and shear
viscous coefficients respectively. The parameters $\alpha_0$ and $\alpha_1$
are connected with the characteristics of the fluid under study. The spatial
projector tensor is designed by $h_{\mu\nu}=g_{\mu\nu}-U_\mu U_\nu$, and 
$\sigma_{\mu\nu}=U_{(\mu;\nu)}-\stackrel{\cdot}{U}_{(\mu}U_{\nu)}-
h_{\mu\nu}U^\mu_{;\mu}/3$ is the shear tensor. Here, the upper dot denotes
$\stackrel{\cdot}{A}_{\alpha\beta...}=U^\lambda 
A_{\alpha\beta...;\lambda}$. 
It can be seen that if some reasonable assumptions are met, these equations 
preserves
the causality condition and predict stable equilibrium configurations \cite
{HiLi83}. However, the transport equations for the dissipative flows
predicted by EIT are extremely involved. Nevertheless, in the linear
approximation the evolution equations for dissipative flows reduces to

$$
\tau _{{\cal Q}}\stackrel{\cdot }{{\cal Q}^\nu }h_\nu^\mu+
{\cal Q}^\mu \simeq \chi h^{\mu \nu }
\left[ T_{,\nu }-T\stackrel{\cdot }{U}_\nu \right] , 
$$
$$
\tau _\Pi U^{\alpha}\Pi_{,\alpha}+\Pi \simeq -\zeta U_{;\mu }^\mu  
$$
and%
$$
\tau _\pi \stackrel{\cdot }{\pi }_{\mu \nu }+\pi _{\mu \nu }\simeq 2\eta
\sigma _{\mu \nu }, 
$$
which are the covariant formulation of Maxwell-Cattaneo transport equations 
\cite{TrPa95,PrHeEs96}. One may hope that the propagation speed of the
dissipative signals is comparable to the sound velocity. For this to be true
the relaxation time in the transport equations must be nearly the radiation
mean free time. After calculating the temperature we shall obtain both, the
bulk and shear viscous pressures, which cannot be directly found from
Einstein equations. In this way we will establish the influence of shear
viscous pressure on the anisotropy of the system.

The paper is organized as follows. In section II the Einstein equations, for
an anisotropic sphere composed by a mixture of radiation and matter, are
derived, and we resort to the HJR method to solve them. In section III is we
write the Maxwell-Cattaneo transport equations within the HJR formalism and
find the mean free path of the neutrinos as a function of the temperature.
This is applied to a specific case in section IV. As initial configuration,
we shall adopt the anisotropic static solution of Gokhroo and Mehra \cite
{GoMe94}. This one corresponds to an anisotropic fluid with variable energy
density. Lastly, section V summarizes the results of this work.

We adopt metrics of signature $-2$. The quantities subscripted with `$a$'
denote that they are evaluated at the surface of the star. The subscripts $0$
and $1$ indicate partial differentiation with respect to time ($u$) and
radial coordinate ($r$), respectively, finally an upper dot on scalar 
quantities means $d/du$. A
hat over a quantity means that this one is measured by a Minkowskian
observer comoving with the fluid.

\section{First set of equations: Einstein field equations}

\subsection{Interior and exterior metrics}

Our aim is to describe a non-static spherically symmetric fluid
distribution. To this end we adopt the radiation coordinates \cite
{BoBuMe62,Bondi64}, then the interior metric takes the form 
\begin{equation}
\label{mbon}ds^2=e^{2\beta }\left[ \frac Vrdu^2+2dudr\right] -r^2\left(
d\theta ^2+sin^2\theta d\phi ^2\right) ,
\end{equation}
where $u~=~x^0$ is a timelike coordinate, $r~=~x^1$ is the null coordinate
and $\theta ~=~x^2$ and $\phi ~=~x^3$ are the usual angle coordinates. The $u
$-coordinate is the retarded time in flat space-time and, therefore, the $u$%
-constant surfaces are null cones open to the future. The metric functions $%
\beta $ and $V$ in equation (\ref{mbon}), are functions of $u$ and $r$. A
function $\tilde m(u,r)$ can be defined by 
\begin{equation}
\label{betav}V=e^{2\beta }(r-2\tilde m(u,r)),
\end{equation}
which is the generalization, inside the distribution, of the ``mass aspect''
defined by Bondi et al \cite{BoBuMe62}. In the static limit it coincides
with the Schwarzschild mass. 

In order to give a clear physical meaning to the above formulae, we
introduce local Minkowskian coordinates $(t,x,y,z)$ related to Bondi's
radiation coordinates by 
\begin{equation}
\label{ccoor}dt=e^\beta (\sqrt{\frac Vr}du+\sqrt{\frac rV}dr),
\end{equation}
\begin{equation}
\label{xccor}dx=e^\beta \sqrt{\frac rV}dr,
\end{equation}
\begin{equation}
\label{ycoor}dy=rd\theta ,
\end{equation}
\begin{equation}
\label{zcoor}dz=rsin\theta d\phi ,
\end{equation}
and to the Schwarzschild coordinates ($T,R,\Theta ,\Phi $) by 
\begin{equation}
\label{tiempsch}T=u+\int_0^r\frac rVdr,
\end{equation}
\begin{equation}
R=r,\ \ \ \Theta =\theta ,\ \ \ \Phi =\phi .
\end{equation}

Outside matter the metric is the Vaidya one \cite{Vaidya51}, a particular
case of the Bondi metric in which $\beta =0$ and $V=r-2m.$

\subsection{Stress-energy tensor}

As we mentioned above we consider an anisotropic fluid sphere composed by a
material medium plus radiation. The material medium travels in the radial
direction with a velocity $\omega $ in a Minkowski coordinate system. The
radiation is treated in the diffusive regime because it interacts strongly
with matter. Due to this interaction, viscous processes appear. Thus, a
viscous term must be considered in the stress-energy tensor. We denote this
term by $\hat T_{\mu \nu }^V$ and, for a local Minkowskian observer comoving
with the fluid, it can be written as 
\begin{equation}
\label{tmunuv}\hat T_{\mu \nu }^V=\hat \tau _{\mu \nu }=\hat \pi _{\mu \nu
}+\Pi \widehat{h}_{\mu \nu }, 
\end{equation}
where $\hat \pi _{\mu \nu }$ denotes the traceless viscous pressure tensor, $%
\widehat{h}_{\mu \nu }~=~\eta _{\mu \nu }~-~\hat U_\mu \hat U_\nu $ the
spatial projection tensor and $\Pi $ is the bulk viscous pressure. For this
comoving observer $\widehat{U}_\mu $ complies with $\widehat{U}_\mu =\delta
_\mu ^t$, and applying the condition $\widehat{\pi }_{\mu \nu }\widehat{U}%
^\mu =0$ together with traceless character of $\hat \pi _{\mu \nu },$ we
obtain 
\begin{equation}
\label{tmunuv2}\widehat{\pi }_{\mu \nu }=\left( 
\begin{array}{cccc}
0 & 0 & 0 & 0 \\ 
0 & \pi & 0 & 0 \\ 
0 & 0 & -\left( \pi /2\right) & 0 \\ 
0 & 0 & 0 & -\left( \pi /2\right) 
\end{array}
\right) . 
\end{equation}

Not all anisotropies can be explained in terms of the interaction between
matter and radiation. Therefore, it is necessary to bear in mind that the
material part of the stress-energy tensor must reflect the anisotropic
character of the fluid. For the Minkowskian observer the anisotropic
material part of the stress-energy tensor $\hat T_{\mu \nu }^M$ is given by
the expression 
\begin{equation}
\label{tmunum}\hat T_{\mu \nu }^M=(\rho _M+P_t)\hat U_\mu \hat U_\nu
-P_t\eta _{\mu \nu }+\left( P-P_t\right) \widehat{\chi }_\mu \widehat{\chi }%
_\nu , 
\end{equation}
where $\widehat{\chi }_\mu =\delta _\mu ^x,$ $\rho _M$ denotes the material
energy density, $P$ the material pressure and $P_t=P+\wp $ the material part
of the tangential pressure. Thus, the pressure $\wp $ refers to the
anisotropies that cannot be explained as a result of the matter-radiation
interaction.

The last term in the stress-energy tensor accounts for the presence of
radiation. In the {\em Lagrangean frame} (the {\em proper frame}) and in the
diffusive regime it reads \cite{Lindquist66,MiMi84} 
\begin{equation}
\label{tmunur}\hat T_{\mu \nu }^R=\left( 
\begin{array}{cccc}
\rho _R & -{\cal Q} & 0 & 0 \\ 
-{\cal Q} & {\cal P} & 0 & 0 \\ 
0 & 0 & {\cal P} & 0 \\ 
0 & 0 & 0 & {\cal P} 
\end{array}
\right) , 
\end{equation}
where $\rho _R$ denotes the radiation energy density, ${\cal Q}$ the heat
flow and ${\cal P}$ the radiation pressure.

Thus, for a local observer comoving with matter, with radial velocity $%
\omega ,$ the stress-energy tensor in local Minkowskian coordinates is%
$$
\hat T_{\mu \nu }=\hat T_{\mu \nu }^M+\hat T_{\mu \nu }^R+\hat T_{\mu \nu
}^V, 
$$
and in virtue of (\ref{tmunuv}), (\ref{tmunuv2}), (\ref{tmunum}) and (\ref
{tmunur}) it can be written as 
\begin{equation}
\label{tmunumi}\hat T_{\mu \nu }=(\rho +P_{\perp })\hat U_\mu \hat U_\nu
-P_{\perp }\eta _{\mu \nu }+(P_r-P_{\perp })\hat \chi _\mu \hat \chi _\nu +2%
\hat {{\cal Q}}_{(\mu }\hat U_{\nu )}, 
\end{equation}
with 
\begin{equation}
\label{usomb}\hat U_\mu =(1,0,0,0), 
\end{equation}
\begin{equation}
\label{chisomb}\hat \chi _\mu =(0,1,0,0), 
\end{equation}
\begin{equation}
\label{qsomb}\hat {{\cal Q}}_\mu ={\cal Q}(0,-1,0,0), 
\end{equation}
\begin{equation}
\rho =\rho _M+\rho _{R,} 
\end{equation}
\begin{equation}
\label{prad}P_r=P+{\cal {P}}+\Pi +\pi , 
\end{equation}
and 
\begin{equation}
\label{ptan}P_{\perp }=P_r-\frac 32\pi +\wp . 
\end{equation}
The physical variables (energy density $\rho $, radial pressure $P_r$,
tangential pressure $P_{\perp }$ and heat flow ${\cal Q}$) are obtained as
measured by the mentioned observer, and the effects of gravitation are
clearly provided through the appropriate transformation to a curvilinear
coordinate system.

To study the dynamics of the system (i.e., $\omega $) it is necessary to
obtain the stress-energy tensor as seen by an observer at rest with respect
to the Minkowskian coordinates. To this end it is necessary to apply a
Lorentz boost in the radial direction to $\hat T^{\mu \nu }.$ A further
transformation allows us to express the stress-energy tensor in curvilinear
coordinates (see \cite{MaPaNu94} for details). Thus, applying a local
coordinate transformation (\ref{ccoor}-\ref{zcoor}) and a Lorentz boost in
the radial direction to (\ref{tmunumi}) we obtain the stress-energy tensor, 
\begin{equation}
\label{ten}T_{\mu \nu }=(\rho +P_{\perp })U_\mu U_\nu -P_{\perp }g_{\mu \nu
}+(P_r-P_{\perp })\chi _\mu \chi \nu +2{\cal Q}_{(\mu }U_{\nu )}, 
\end{equation}
as measured by an observer using Bondi coordinates with a radial velocity,
with respect to the matter configuration, $-\omega .$ After performing the
transformation, expressions (\ref{usomb}-\ref{qsomb}) reads 
\begin{equation}
\label{velo}U_\mu =e^\beta \left( 
\begin{array}{c}
\begin{array}{cccc}
\sqrt{\frac Vr}\frac 1{(1-\omega ^2)^{1/2}}, & \sqrt{\frac rV}\left( \frac{%
1-\omega }{1+\omega }\right) ^{1/2}, & 0, & 0 
\end{array}
\end{array}
\right) , 
\end{equation}
\begin{equation}
\chi _\mu =e^\beta \left( 
\begin{array}{cccc}
-\sqrt{\frac Vr}\frac \omega {(1-\omega ^2)^{1/2}}, & \sqrt{\frac rV}\left( 
\frac{1-\omega }{1+\omega }\right) ^{1/2}, & 0, & 0 
\end{array}
\right) , 
\end{equation}
and 
\begin{equation}
\label{q}{\cal Q}_\mu =-{\cal Q}\chi _\mu . 
\end{equation}
Note that ${\cal Q}^\mu U_\mu =0$ and ${\cal Q=}\sqrt{-{\cal Q}^\mu {\cal Q}%
_\mu }.$

Applying these transformations to the traceless viscous tensor (\ref{tmunuv2}%
), we get 
\begin{equation}
\label{pimunufinal}\pi _{\mu \nu }=\left( 
\begin{array}{cccc}
e^{2\beta }\frac Vr\left( \frac{\omega ^2}{1-\omega ^2}\right) \pi & 
-e^{2\beta }\left( \frac \omega {1+\omega }\right) \pi & 0 & 0 \\ 
-e^{2\beta }\left( \frac \omega {1+\omega }\right) \pi & e^{2\beta }\frac rV%
\left( \frac{1-\omega }{1+\omega }\right) \pi & 0 & 0 \\ 
0 & 0 & -\frac{r^2}2\pi & 0 \\ 
0 & 0 & 0 & -\frac{r^2\sin ^2\theta }2\pi 
\end{array}
\right) . 
\end{equation}

Outside matter the stress-energy tensor corresponds to that of a null fluid,
i.e. 
\begin{equation}
T^{\mu \nu }=\varepsilon k^\mu k^\nu , 
\end{equation}
where 
\begin{equation}
k^\mu =\delta _r^\mu e^{-2\beta }\sqrt{\frac Vr}. 
\end{equation}

\subsection{Einstein field equations}

Inside matter the Einstein field equations, $G_{\mu \nu }=8\pi T_{\mu \nu },$
can be written as:

\begin{equation}
\label{ecu00}\frac 1{4\pi r(r-2\tilde m)}\left( -\tilde m_0e^{-2\beta }+(1-2%
\tilde m/r)\tilde m_1\right) =\frac 1{1-\omega ^2}(\rho +2\omega {\cal Q}%
+P_r\omega ^2), 
\end{equation}
\begin{equation}
\label{ecu01}\frac{\tilde m_1}{4\pi r^2}=\frac 1{1+\omega }(\rho -{\cal Q}%
(1-\omega )-P_r\omega ), 
\end{equation}
\begin{equation}
\label{ecu11}\beta _1\frac{r-2\tilde m}{2\pi r^2}=\frac{1-\omega }{1+\omega }%
(\rho -2{\cal Q}+P_r), 
\end{equation}
\begin{equation}
\label{ecu22}-\frac{\beta _{01}e^{-2\beta }}{4\pi }+\frac 1{8\pi }(1-2\frac{%
\tilde m}r)(2\beta _{11}+4\beta _1^2-\frac{\beta _1}r)+\frac{3\beta _1(1-2%
\tilde m_1)-\tilde m_{11}}{8\pi r}=P_{\perp }, 
\end{equation}
while outside matter, the only non vanishing Einstein equation yields 
\begin{equation}
\widetilde{m}_0=-4\pi r^2\varepsilon \left( 1-\frac{2\widetilde{m}(u)}r%
\right) . 
\end{equation}

To algebraically solve the physical variables present in the above set of
field equations (\ref{ecu00}-\ref{ecu22}) we must find out both $\beta $ and 
$\tilde m.$ The HJR method will help us in this task, but before it we must
impose the junction conditions between exterior and interior metrics.

\subsection{Junction conditions and surface equations}

The mass function can be expressed as 
\begin{equation}
\label{masa}\tilde m=\int_0^r4\pi r^2\tilde \rho dr. 
\end{equation}
It involves an effective energy density given by the right-hand side of (\ref
{ecu01}), 
\begin{equation}
\label{rotilde}\tilde \rho =\frac 1{1+\omega }(\rho -{\cal Q}(1-\omega
)-P_r\omega ), 
\end{equation}
which in the static limit reduces to the energy density of the system.

From (\ref{ecu11}) one has 
\begin{equation}
\beta =\int_{a(u)}^r\frac{2\pi r^2}{r-2\tilde m}\frac{1-\omega }{1+\omega }%
(\rho -2{\cal Q}+P_r)dr, 
\end{equation}
and we may re-write the non-static case as 
\begin{equation}
\label{betatil}\beta =\int_{a(u)}^r\frac{2\pi r^2}{r-2\tilde m}(\tilde \rho +%
\tilde P)dr, 
\end{equation}
with 
\begin{equation}
\label{ptilde}\tilde P=\frac 1{1+\omega }(-\omega \rho -{\cal Q}(1-\omega
)+P_r) 
\end{equation}
being the effective pressure, which also reduces to the radial pressure in
the static limit.

Matching the Vaidya metric to the Bondi metric at the surface ( $r=a$) of
the fluid distribution implies $\beta _a=\beta(u,r=a)=0$ with the continuity
of the mass function $\tilde m(u,r)$ (i.e. the continuity of the first
fundamental form) and the continuity of the second fundamental form leads to

\begin{equation}
\label{jc3}\dot a=-\left( 1-2\frac{\tilde m_a}a\right) \frac{\tilde P_a}{%
\tilde P_a+\tilde \rho _a}. 
\end{equation}
(see \cite{HeJi83} for details).

From the coordinate transformation (\ref{ccoor}) the velocity of matter in
Bondi coordinates can be written as 
\begin{equation}
\label{drdu}\frac{dr}{du}=\frac Vr\frac \omega {1-\omega }, 
\end{equation}
evaluating the last expression at the surface and comparing it with (\ref
{jc3}) it follows 
\begin{equation}
\label{jc5}\tilde P_a=-\omega _a\tilde \rho _a, 
\end{equation}
or equivalently using (\ref{rotilde}) and (\ref{ptilde}) we get 
\begin{equation}
\label{jc6}{\cal Q}_a=P_{ra}, 
\end{equation}
which is a well known result for radiative spheres \cite{Santos85}.

To derive the surface equations we introduce five dimensionless functions 
\begin{equation}
\label{adime}A\equiv \frac a{\widetilde{m}(0)}\;\;\;\;M\equiv \frac{%
\widetilde{m}}{\widetilde{m}(0)}\;\;\;\;u\equiv \frac u{\widetilde{m}(0)}%
\;\;\;\;F\equiv 1-\frac{2M}A\;\;\;\;\Omega \equiv \frac 1{1-\omega _a}, 
\end{equation}
where $\widetilde{m}(0)$ is the initial mass of the system. Using the
functions just defined into (\ref{jc3}) we get the first surface equation 
\begin{equation}
\label{se1}\dot A=F(\Omega -1). 
\end{equation}
This first surface equation gives the evolution of the radius of the star.

The second surface equation emerges from the luminosity evaluated at the
surface of the system. The luminosity as seen by a comoving observer is
defined as 
\begin{equation}
\label{ehat}\hat E=(4\pi r^2{\cal Q})_{r=a}. 
\end{equation}
Evaluating (\ref{ecu00}) and (\ref{ecu01}) at the surface and using the
expansion 
\begin{equation}
\label{m0a}\tilde m_{0_a}\approx \tilde m-\dot a\tilde m_{1_a}, 
\end{equation}
the luminosity perceived by an observer at rest at infinity reads 
\begin{equation}
\label{luminosity}L=-\dot M=\hat E(2\Omega -1)F=4\pi A^2{\cal Q}_a\left(
2\Omega -1\right) F. 
\end{equation}
The function $F$ is related to the boundary redshift $z_a$ by 
\begin{equation}
1+z_a=\frac{\nu _{em}}{\nu _{rec}}=F^{-1/2}. 
\end{equation}
Thus the luminosity as measured by a non comoving observer located on the
surface is 
\begin{equation}
\label{esinhat}E=L(1+z_a)^2=-\frac{\dot M}F=\hat E(2\Omega -1), 
\end{equation}
where the term $(2\Omega -1)$ accounts for the boundary doppler-shift. Using
relationship (\ref{luminosity}) together with the first surface equation we
obtain the second one as 
\begin{equation}
\label{se2}\dot F=\frac{2L+F(1-F)(\Omega -1)}A, 
\end{equation}
which express the evolution of the redshift at the surface.

The third surface equation is model dependent. For anisotropic fluids the
relationship $(T_{r;\mu }^\mu )_a=0$ can be written as 
\begin{equation}
\label{tov}-\left( \frac{\tilde P+\tilde \rho }{1-2\tilde m/r}\right) _{0a}+%
\tilde R_{\perp _a}-\left( \frac 2r(P_r-\tilde P)\right) _a=0, 
\end{equation}
where 
\begin{equation}
\tilde R_{\perp _a}=\tilde P_{1_a}+\left( \frac{\tilde P+\tilde \rho }{1-2%
\tilde m/r}\right) _a\left( 4\pi r\tilde P+\frac{\tilde m}{r^2}\right)
_a-\left( \frac 2r(P_{\perp }-P_r)\right) _a. 
\end{equation}
It is possible to associate a physical meaning to $\tilde R_{\perp a}$: the
first term is tied to the hydrodynamic force, the second one to the
gravitational force and the third one reflects the anisotropic character of
the fluid. The latter is negative when the radial pressure is larger than
the tangential pressure. In this case the fluid distribution may become more
compact than in the isotropic one.

Using the expansions (\ref{m0a}) and 
$$
(\tilde \rho +\tilde P)_{0_a}\approx [\tilde \rho _a(1-\omega _a)]_0-\dot a(%
\tilde P+\tilde \rho )_{1_a}, 
$$
where (\ref{jc5}) has been used, we get 
$$
\frac{\dot F}F+\frac{\dot \Omega }\Omega -\frac{\tilde \rho }{\tilde \rho _a}%
+F\Omega ^2\frac{\tilde R_{\perp _a}}{\tilde \rho _a}-\frac 2AF\Omega \frac{%
P_{ra}}{\tilde \rho _a}= 
$$
\begin{equation}
\label{se3}(1-\Omega )\left[ 4\pi A\tilde \rho _a\frac{3\Omega -1}\Omega -%
\frac{3+F}{2A}+F\Omega \frac{\tilde \rho _{1_a}}{\tilde \rho _a}+\frac{%
2F\Omega }{A\tilde \rho _a}(P_{\perp }-P_r)_a\right] . 
\end{equation}
Expression (\ref{se3}) generalizes the Tolman-Oppenheimer-Volkov equation to
the non-static radiative anisotropic case.

Now we are in position to brief summarize the HJR method \cite{HeJiRu80}
applied to the anisotropic case.

\subsection{The HJR method}

This method has been extensively used since last decade to obtain models of
non static radiating fluids from known static solutions of the Einstein
equations -\cite{MaPaNu94,HeJiRu80,HeNu90} and references therein. Its use
reduces the problem of the gravitational collapse to the solution of the
surface equations (\ref{se1}-\ref{se3}). This procedure simplifies the
resolution of the problem because these ones are a set of ordinary
differential equations that can be solved by Runge-Kutta method. The
algorithm can be summarized as follows.

\begin{enumerate}
\item  Take a static but otherwise arbitrary interior solution of the
Einstein equations for a spherically symmetric fluid distribution 
$$
P_{st}=P(r),\ \ \ \ \ \rho _{st}=\rho (r). 
$$

\item  The effective quantities $\tilde \rho \equiv \tilde \rho (u,r)$ and $%
\tilde P\equiv \tilde P(u,r)$ must coincide with $\rho _{st}$ and $P_{st}$
respectively in the static limit. We assume that the $r$-dependence in
effective quantities is the same that in its corresponding static ones.
Nevertheless, in non static case the junction conditions are different
accomplishing expression (\ref{jc5}). This condition allows us to find out
the relation between the $u$-dependence of $\tilde \rho \equiv \tilde \rho
(u,r)$ and $\tilde P\equiv \tilde P(u,r)$.

\item  Introduce $\tilde \rho (u,r)$ and $\tilde P(u,r),$ into (\ref{masa} )
and (\ref{betatil}) to determine $\tilde m$ and $\beta $ up to three unknown
functions of time.

\item  The three surface equations form a system of first order ordinary
differential equations, by solving it we find the evolution of the radius, $%
A(u)$, and two unknown functions of time. These ones can be related with the 
$u$-dependence of $\tilde \rho \equiv \tilde \rho (u,r)$ and $\tilde P\equiv 
\tilde P(u,r)$.

\item  There are four unknown functions of time ($A$, $F$, $\Omega $ and $L$%
). Thus, it is necessary to impose the evolution of one of them to solve the
system of three surface equations. Usually the luminosity is taken as an
input function because it can be found from observational quantities.

\item  Once these three functions are known, it is easy to find $\tilde m$
and $\beta .$ Therefore, the interior metric is completely defined.

\item  Now, the left-hand side of the Einstein equations (\ref{ecu00}-\ref
{ecu22}) is known. However, the right-hand side of these equations contain
five unknown quantities ($\omega $, $\rho $, $P_r$, $P_{\perp }$ and ${\cal Q%
}$). Thus, it is necessary to supply another equation to close the system of
field equations. In the anisotropic static case it can be found a general
equation that relates the tangential pressure to the mass function, energy
density and radial pressure \cite{CoHeEs81}, 
\begin{equation}
P_{\perp }-P_r=\frac r2P_1+\left( \frac{\rho +P}2\right) \left( \frac{m+4\pi
r^3P}{r-2m}\right) .
\end{equation}
This expression is usually generalized, in the context of HJR method, to
non-static cases by substituting for the effective variables the physical
quantities \cite{BaRo92,CoHeEsWi82} 
\begin{equation}
\label{ptmpr}P_{\perp }-P_r=\frac r2\widetilde{P}_1+\left( \frac{\widetilde{%
\rho }+\widetilde{P}}2\right) \left( \frac{\widetilde{m}+4\pi r^3\widetilde{P%
}}{r-2\widetilde{m}}\right) .
\end{equation}
Now, the Einstein equations, supplemented with (\ref{ptmpr}), make up a
close system of equations and quantities $\omega $, $\rho $, $P_r$, $%
P_{\perp }$ and ${\cal Q}$ can be found.
\end{enumerate}

The HJR method starts from a static interior solution of the Einstein
equations ( $P_{st}=P(r)$ and $\rho _{st}=\rho (r)$). The effective
variables (\ref{rotilde}) and (\ref{ptilde}) must reduce in the static case
to $\rho _{st}$ and $P_{st}$ respectively. Nevertheless, these quantities
have a dependence on the retarded time $u.$ We assume that the effective
variables have the same $r$-dependence as the physical variables of the
static situation have. This can be justified in terms of the characteristic
times for different processes involved in the collapse scenario. If the
hydrostatic time scale ${\cal T}_{HYDR}$, $\sim 1/\sqrt{G\rho },$ is much
shorter than the Kelvin-Helmholtz time scale, ${\cal T}_{KH}$, then in first
approximation the inertial terms in the equation of motion $\left[ T_{r;\mu
}^\mu \right] _{r=a}=0$ can be ignored \cite[page 11]{KiWe90}. The
Kelvin-Helmholtz phase of the birth of a neutron star last some tens of
seconds \cite{BuLa86}, whereas for a neutron star of one solar mass and a
ten-kilometer radius, we obtain ${\cal T}_{HYDR}\sim 10^{-4}$s. Thus, in
this first approximation the $r$-dependence of physical quantities $P$ and $%
\rho $ are the same as in the static solution. The assumption that not the
physical quantities but the effective variables (\ref{rotilde}) and (\ref
{ptilde}) have the same $r$-dependence as the physical variables of the
static situation, is a correction to that first approximation. Therefore it
is expected that the introduction of two functions of time in $\widetilde{%
\rho }$ and $\widetilde{P}$ preserving the same $r$-dependence as in $\rho
_{st}$ and $P_{st}$ will yield good results. These two functions of time can
be related by means of the junction condition (\ref{jc5}).

If the evolution of effective variables where known the functions $%
\widetilde{m}$ and $\beta $ could be found and the Einstein equations
supplemented by (\ref{ptmpr}) would constitute a closed system of
differential equations. This method will be clarified by means of an example
in section IV.

\section{Second set of equations: Transport equations}

We shall use the method described above to find the heat flow, the energy
density, the radial and tangential pressures and the velocity of the
collapse. However, if one wished to explicitly find the temperature and
viscous pressure, one has to resort to the transport equations laid down by
EIT.

Usually classical theory \cite{Eckart40,KlGrMa53,LaLi71} has been employed
as a first approximation to the study of gravitational collapse.
Nevertheless, this one presents two important disabilities \cite{HiLi83}:
(1) It predicts an infinite speed for thermal and viscous signals. (2) The
equilibrium states turn out to be unstable. Therefore, as a further step in
the thermodynamic study of gravitational collapse, it seems suitable to
resort to a theory free of such drawbacks. As we pointed out in section I
the relativistic EIT may do the job. The Maxwell-Cattaneo transport
equations are a particular case of those given by EIT -see for instance \cite
{TrPa95,PrHeEs96}. The former are decoupled expressions of the EIT transport
equations and they may be understood as a first approximation to them. Thus,
we consider that the dissipative flows are decoupled and the transport
equations may approximated by their Maxwell-Cattaneo form. The heat
conduction equation can be written as 
\begin{equation}
\label{flujo}\tau _{{\cal Q}}\stackrel{\cdot }{{\cal Q}^\nu }h_\nu^\mu+
{\cal Q}^\mu
\simeq \chi h^{\mu \nu }\left[ T_{,\nu }-T\stackrel{\cdot }{U}_\nu \right] ,
\end{equation}
where $h^{\mu \nu }=g^{\mu \nu }-U^\mu U^\nu $ is the spatial projection
tensor, $\chi $ is the thermal conductivity coefficient, $T$ the temperature
and $\tau _{{\cal Q}}$ is the relaxation time for thermal signals. The
evolution of bulk viscous pressure $\Pi $ is given by the expression 
\begin{equation}
\label{bulk}\tau _\Pi U^{\alpha}\Pi_{,\alpha}+\Pi \simeq -\zeta U_{;\mu
}^\mu .
\end{equation}
The bulk viscosity coefficient and relaxation time for bulk viscous signals
are denoted by $\zeta $ and $\tau _\Pi $ respectively. On the other hand the
evolution of shear viscous pressure $\pi _{\mu \nu }$ (\ref{pimunufinal})
can be written in Maxwell-Cattaneo form as, 
\begin{equation}
\label{tang}\tau _\pi \stackrel{\cdot }{\pi }_{\mu \nu }+\pi _{\mu \nu
}\simeq 2\eta \sigma _{\mu \nu },
\end{equation}
where shear tensor $h_{(\mu }^\alpha U_{\nu );\alpha }-U_{;\alpha }^\alpha
h_{\mu \nu }/3$ is denoted by $\sigma _{\mu \nu }$, $\tau _\pi $ corresponds
to relaxation time for shear viscous signals and $\eta $ indicates the shear
viscosity coefficient.

Eckart transport equations are given by 
\begin{equation}
\label{eck}{\cal Q}^\mu \simeq \chi h^{\mu \nu }\left[ T_{,\nu }-T\stackrel{%
\cdot }{U}_\nu \right] , 
\end{equation}
\begin{equation}
\Pi \simeq -\zeta U_{;\mu }^\mu 
\end{equation}
and 
\begin{equation}
\pi _{\mu \nu }\simeq 2\eta \sigma _{\mu \nu }. 
\end{equation}
Note that the main difference between both formulations is the presence of
the relaxation time in (\ref{flujo}-\ref{tang}). As we will show in section
IV this discordance is important close to surface.

To solve this system of partial differential equations it is necessary to
adopt an expression for transport coefficients $\chi $, $\eta $ and $\zeta $%
. For a mixture of matter and radiation these ones are given by (see e.g{\it %
.} \cite{Weinberg71}) 
\begin{equation}
\label{chicoeff}\chi =\frac 43bT^3\tau , 
\end{equation}
\begin{equation}
\label{etacoeff}\eta =\frac 4{15}bT^4\tau 
\end{equation}
and 
\begin{equation}
\label{bulkcoeff}\zeta =15\eta \left[ \frac 13-\left( \frac{\partial P}{%
\partial \rho }\right) _n\right] ^2, 
\end{equation}
where $\tau $ denotes the mean free time of radiation (in our case
neutrinos), while the constant $b$ takes (for neutrinos) the value $7a/8$,
where $a$ is radiation constant.

Expressions (\ref{flujo}-\ref{tang}) implies a finite propagation speed for
viscous and thermal signals. Nevertheless, to guarantee relativistic
causality it is necessary to restrict the possible values of the different
relaxation times. If we assume that the propagation speed of thermal and
viscous signals are of the order of sound speed, then 
\begin{equation}
\label{condtau}\tau _{{\cal Q}}\sim \tau _\Pi \sim \tau _\pi \sim \tau . 
\end{equation}
This means that the relaxation times are somewhat larger than the mean free
time of radiation. Thus, if $\tau $ is known, it is possible to use (\ref
{flujo}) to find the temperature and, by (\ref{bulk}) and (\ref{tang}) to
calculate the viscous pressures.

\subsection{Determination of $\tau $ for neutrinos}

Thermal neutrino processes are important during the late stages of collapse
of massive stars \cite{ShTe83}. They help to carry thermal energy from the
core to outer regions. Thus, we assume that neutrinos are principally
thermally generated with energies close to $k_BT$.

A detailed study of the contribution of different processes to the neutrino
transport has been given by Bruenn \cite{Bruenn85}. Our aim is to establish
an approximate expression to the mean free time for neutrinos. Despite the
procedure used in this work yields good results, it is highly idealized and
only approximate. A more accurate expression for $\tau $ would be desirable,
but we consider that the approach below describes the process accurately
enough.

At the typical densities of a neutron star, matter can capture neutrinos.
The neutrino trapping helps to drive the neutrinos to local equilibrium. The
matter opacity to the neutrinos takes place principally by means of
electron-neutrino scattering ($e^{-}+\nu \rightarrow e^{-}+\nu $) and
nucleon absorption ($\nu +n\rightarrow p+e^{-}$).

As a first approximation the cross-section for nucleon absorption may be
taken as \cite[chapter 18]{ShTe83} 
\begin{equation}
\sigma _n\approx 10^{-44}\varepsilon _\nu ^2\text{ cm}^2, 
\end{equation}
where $\varepsilon _\nu $ is the energy of neutrinos in MeV. Then, the
neutrino mean free path is given by 
\begin{equation}
\lambda _n=\frac 1{n_n\sigma _n}\approx \frac{10^{20}}{\rho \varepsilon _\nu
^2}\text{ cm,} 
\end{equation}
where $n_n\approx 6\times 10^{23}\rho $ cm$^{-3}$ \cite{ShTe83,HaKa94}, is
the neutron number density.

In high energetic collisions the cross-section for the electron-neutrino
scattering can be approximated by \cite{HaMa83} 
\begin{equation}
\sigma _e\approx 10^{-44}\varepsilon _\nu \text{ cm}^2. 
\end{equation}
The electron number density $n_e$, can be related to $n_n$ by means of the
electron fraction ${\cal Y}_e$ ($n_e\simeq n_n{\cal Y}_e$). Hence, the mean
free path is 
\begin{equation}
\lambda _e=\frac 1{n_e\sigma _e}\approx \frac{10^{20}}{\rho \varepsilon _\nu 
{\cal Y}_e}\text{ cm.} 
\end{equation}

The effective mean free time which takes into account the neutrino zig-zag
path may be written as 
\begin{equation}
\label{noseque}\tau =\lambda \sim \sqrt{\lambda _e\lambda _n}\approx \frac{%
10^{20}}{\rho \sqrt{{\cal Y}_e\varepsilon _\nu ^3}}\text{ cm.} 
\end{equation}

Assuming that the neutrinos are generated by thermal emission ($\varepsilon
_\nu \sim k_BT$) we find the $\tau $ dependence on temperature 
\begin{equation}
\tau \propto T^{-3/2}. 
\end{equation}
A rigorous treatment of mean free time would comprise the evolution of $%
{\cal Y}_e$. Nevertheless, this effort doesn't seem advisable here due to
the degree of the approximation adopted and that the range of possible
values for ${\cal Y}_e$ is severely restricted ($0.2\leq {\cal Y}_e\leq 0.3$%
) \cite{Bruenn85}.

It is convenient to express $\tau $ in dimensionless form to simplify the
numerical treatment of equations (\ref{flujo}-\ref{tang}). In virtue of (\ref
{adime}) and (\ref{noseque}) 
\begin{equation}
\label{time}\tau \sim {\cal A}\frac{M_o}{\rho \sqrt{{\cal Y}_eT^3}}, 
\end{equation}
where the constant ${\cal A}$ takes the value $10^9$ K$^{3/2}$ m$^{-1},$ $%
M_o $ is the initial mass of star in meters, $T$ the Kelvin temperature and $%
\rho $ the dimensionless energy density.

\subsection{Maxwell-Cattaneo transport equations: explicit form}

The Maxwell-Cattaneo transport equations (\ref{flujo}-\ref{tang}) can be
written, by means of (\ref{chicoeff}-\ref{bulkcoeff}) and condition (\ref
{condtau}), in dimensionless form 
\begin{equation}
\tau \stackrel{\cdot }{{\cal Q}^\nu }h_\nu^\mu+{\cal Q}^\mu \simeq \frac 76%
aM_o^2T^3\tau h^{\mu \nu }\left[ T_{,\nu }-T\stackrel{\cdot }{U}_\nu \right]
, 
\end{equation}
\begin{equation}
\tau U^{\alpha}\Pi_{,\alpha}+\Pi \simeq -\frac 72aM_o^2T^4\tau \left[ \frac 1%
3-\left( \frac{\partial P}{\partial \rho }\right) _n\right] ^2U_{;\mu }^\mu
, 
\end{equation}
\begin{equation}
\tau \stackrel{\cdot }{\pi }_{\mu \nu }+\pi _{\mu \nu }\simeq \frac 7{15}%
aM_o^2T^4\tau \sigma _{\mu \nu }, 
\end{equation}
where $a\simeq 6.252\times 10^{-64}$ cm$^{-2}$ K$^{-4}.$

The expression of mean free time for neutrinos (\ref{time}) is given for a
comoving observer. Thus, applying a Lorentz boost to it, it is possible to
write the transport equations for an observer with radial velocity $-\omega $
respect to matter, in Bondi coordinates.

As we saw in last section, by means of the method HJR it is possible to
solve the Einstein equations. Thus, the quantities ${\cal Q}$, $\rho $, $P_r$%
, $P_{\perp }$ and $\omega $ and the metric functions $\beta $ and $%
\widetilde{m}$ are known. Resorting to expressions (\ref{time}), (\ref{mbon}%
), (\ref{velo}) and (\ref{q}) it is possible, after a
straightforward calculation, to write down the Maxwell-Cattaneo equations in
terms of known quantities ${\cal Q},\ \widetilde{m},\ \beta ,$ $\omega ,\
\rho $ and $P_r$ and their derivatives.

The evolution of the heat flow (\ref{flujo}) is governed by 
\begin{equation}
\label{difeflujo}f(u,r)\left( \frac{aM_o^2}\phi \right) ^{3/5}+\frac{g(u,r)}{%
{\cal B}}\simeq \phi _0+h(u,r)\phi _1+w(u,r)\phi , 
\end{equation}
where we have introduced for sake of simplicity 
\begin{equation}
\phi =aM_o^2T^{5/2}, 
\end{equation}
$$
f(u,r)=\frac{15}7\left[ \frac{{\cal Q}\omega }{(1-\omega ^2)}\left[ \omega
_0+\frac Vr\left( \frac \omega {1-\omega }\right) \omega _1\right] +\left[ 
{\cal Q}_0+\frac Vr\left( \frac \omega {1-\omega }\right) {\cal Q}_1\right]
\right] 
$$
$$
-\frac{15}7\frac{e^{2\beta }{\cal Q}\omega }{1-\omega }\left[ -2\beta
_1\left( 1-\frac{2\widetilde{m}}r\right) +\frac 1r\left( \widetilde{m}_1-%
\frac{\widetilde{m}}r\right) +\frac{\widetilde{m}_0}V(1-\omega )\right] 
$$
\begin{equation}
+\frac{15}7{\cal Q}\omega\frac{V}{r} \left( \frac{1+\omega}{1-\omega} \right)%
\stackrel{\cdot }{U}_r,
\end{equation}
\begin{equation}
g(u,r)=\frac{15}7{\cal Q}e^{2\beta }\sqrt{1-\frac{2\widetilde{m}}r}(1+\omega
), 
\end{equation}
\begin{equation}
h(u,r)=-\frac{e^{2\beta }}{1-\omega }\left( 1-\frac{2\widetilde{m}}r\right)
, 
\end{equation}
\begin{equation}
w(u,r)=\frac 52e^{2\beta }\left( 1-\frac{2\widetilde{m}}r\right) \left( 
\frac{1+\omega }{1-\omega }\right) \stackrel{\cdot }{U}_r, 
\end{equation}
$$
\stackrel{\cdot }{U}_r=\frac 1{1+\omega }\left[ \frac 1{2r}-2\beta _1-\frac{%
1-2\widetilde{m}_1}{2(r-2\widetilde{m})}\right] +re^{-2\beta }\left( \frac{%
1-\omega }{1+\omega }\right) \frac{\widetilde{m}_0}{(r-2\widetilde{m})^2} 
$$
\begin{equation}
-\frac 1{(1+\omega )^2(1-\omega )}\left[ \omega \omega _1+re^{-2\beta }\frac{%
1-\omega }{r-2\widetilde{m}}\omega _0\right] , 
\end{equation}
and 
\begin{equation}
{\cal B}={\cal A}\frac{M_o}{\rho \sqrt{{\cal Y}_e}}. 
\end{equation}

The bulk viscous pressure (\ref{bulk}) can be found by solving the partial
differential equation 
$$
-\frac 72aM_o^2T^4e^{2\beta}\sqrt{1-2\frac{\widetilde{m}}{r}}
\sqrt{\frac{1+\omega}{1-\omega}}\left[ \frac 13-\left( \frac{\partial P}{\partial \rho }%
\right) _n\right] ^2\theta \simeq \Pi _0 
$$
\begin{equation}
\label{bdifepvv}+\left[ e^{2\beta }\left( 1-\frac{2\widetilde{m}}r\right) 
\frac \omega {1-\omega }\right] \Pi _1+\left[ e^{2\beta }\sqrt{1-\frac{2%
\widetilde{m}}r}\left( 1+\omega \right) \frac{T^{3/2}}{{\cal B}}\right]\Pi. 
\end{equation}
where the expansion $\theta $ can be written as%
$$
\theta =U_{;\mu }^\mu =\sqrt{1-\frac{2\widetilde{m}}r}\frac{2\omega }{\sqrt{%
1-\omega ^2}}\left( \frac 1r+\beta _1\right) +\frac 1{\sqrt{1-%
%TCIMACRO{\dfrac{2\widetilde{m}}r }
%BeginExpansion
{\displaystyle {2\widetilde{m} \over r}}
%EndExpansion
}}\frac \omega {\sqrt{1-\omega ^2}}\left( \frac{\widetilde{m}}{r^2}-\frac{%
\widetilde{m}_1}r\right) 
$$
\begin{equation}
+\ e^{-2\beta }\frac{\widetilde{m}_0}{r\left( 1-%
%TCIMACRO{\dfrac{2\widetilde{m}}r }
%BeginExpansion
{\displaystyle {2\widetilde{m} \over r}}
%EndExpansion
\right) ^{3/2}}\sqrt{\frac{1-\omega }{1+\omega }}+\frac{\sqrt{1-%
%TCIMACRO{\dfrac{2\widetilde{m}}r }
%BeginExpansion
{\displaystyle {2\widetilde{m} \over r}}
%EndExpansion
}}{\sqrt{1-\omega ^2}\left( 1+\omega \right) }\left[ \frac{\omega _1}{%
1-\omega }-e^{-2\beta }\frac{\omega _0}{1-%
%TCIMACRO{\dfrac{2\widetilde{m}}r }
%BeginExpansion
{\displaystyle {2\widetilde{m} \over r}}
%EndExpansion
}\right] . 
\end{equation}

The description of shear viscous pressure evolution (\ref{tang}) is given,
as in the other two dissipative flows, by a first order partial differential
equation%
$$
-\frac{14M_o^2}{15\sqrt{3}}aT^4e^{2\beta }\sqrt{1-\frac{2\widetilde{m}}r}%
\sqrt{\frac{1+\omega }{1-\omega }}\sigma \simeq \pi _0 
$$

\begin{equation}
\label{bdifepvt}+\left[ e^{2\beta }\left( 1-\frac{2\widetilde{m}}r\right) 
\frac \omega {1-\omega }\right] \pi _1+\left[ e^{2\beta }\sqrt{1-\frac{2%
\widetilde{m}}r}\left( 1+\omega \right) \frac{T^{3/2}}{{\cal B}}\right] \pi
, 
\end{equation}
where $\sigma ^2=\frac 12\sigma _{\mu \nu }\sigma ^{\mu \nu }$ and, 
\begin{equation}
\sigma =\frac{\sqrt{3}}r\left[ -\frac \omega {\sqrt{r}}\sqrt{\frac{r-2%
\widetilde{m}}{1-\omega ^2}}+\frac \theta 3r\right] . 
\end{equation}

The transport coefficients (\ref{chicoeff}-\ref{bulkcoeff}) as measured by a
comoving observer read, 
\begin{equation}
\label{chi}\chi =\frac 43bc^2T^{3/2}\frac \Upsilon {\rho \sqrt{{\cal Y}_e}}, 
\end{equation}
\begin{equation}
\label{eta}\eta =\frac 4{15}bT^{5/2}\frac \Upsilon {\rho \sqrt{{\cal Y}_e}}, 
\end{equation}
\begin{equation}
\label{gi}\zeta =4bT^{5/2}\left[ \frac 13-\left( \frac{\partial P}{\partial
\rho }\right) _n\right] ^2\left[ \frac \Upsilon {\rho \sqrt{{\cal Y}_e}}%
\right] , 
\end{equation}
with $\Upsilon \simeq 4.173\times 10^{24}$ g cm$^{-3}$ K$^{3/2}$ s.

The system of three partial differential equations (\ref{difeflujo}), (\ref
{bdifepvv}) and (\ref{bdifepvt}) can be solved provided initials, finals and
boundary conditions are given. The boundary condition for temperature in the
heat conduction equation (\ref{difeflujo}) deserves special attention. As
shown in section IV C, the boundary condition for the temperature must
fulfill some requirements. After solving the heat conduction equation, we
shall find the temperature, and using it in the evolution equations (\ref
{bdifepvv},\ref{bdifepvt}) the bulk and shear viscous pressure will be
determined. Once the temperature is known, it will be easy to find the
coefficients of heat conductivity, shear viscosity and bulk viscosity from
equations (\ref{chi}-\ref{gi}), respectively.

\section{Application to the Gokhroo \& Mehra's configuration.}

As we have seen in section III, the HJR method starts from a static solution
of Einstein's equations. It is possible to find numerical models based in
nuclear physics as solutions of the Einstein equations. However, the
uncertainties introduced by this procedure in the HJR method makes it
desirable to start from a static analytical model instead. On the other
hand, it is suitable to adopt an initially static fluid distribution not
excessively idealized. Some analytic anisotropic static solutions of the
field equations are known \cite
{Canuto74,BoLi74,Bayin82,Ponce87a,Ponce87b,Ponce87c,MaMa89}. We shall adopt
the static form of Gokhroo and Mehra (GM) \cite{GoMe94} to illustrate the
application of method described above. This solution corresponds to a fluid
with variable energy density in which some anisotropy is initially present
even in absence of radiation. It leads, under some circumstances, to
densities and pressures similar to the Bethe-B\"orner-Sato{\it \ Newtonian}
state equation (BBS) \cite{ShTe83,Demianski85,KiWe90,Borner73}. Thus, this
solution presents a compromise between the realism of the solutions based in
nuclear physics and the desirable analytic expression of the static solution.

\subsection{Introduction of the GM configuration in the HJR method}

Following the first point of the HJR method (section II E), it is necessary
to start from a static solution ($\rho _{st}$ and $P_{st}$) of the Einstein
equations. In the static solution of Gokhroo and Mehra \cite{GoMe94} the
energy density and radial pressure are assumed to be 
\begin{equation}
\label{statenergy}\rho (r)=\rho _c\left( 1-k\frac{r^2}{a^2}\right) 
\end{equation}
and 
\begin{equation}
\label{statpressure}P_r(r)=P_c\left( 1-\frac{2m}r\right) \left( 1-\frac{r^2}{%
a^2}\right) ^n, 
\end{equation}
where $0\leq k\leq 1$ and $n\geq 1$ are constants. The central energy
density and radial pressure are denoted by $\rho _c$ and $P_c,$
respectively, and in the static case they are related by means of a constant 
$\lambda $ through the expression 
\begin{equation}
\label{pcversusroc}P_c=\lambda \rho _c. 
\end{equation}
Thus, we identify $\rho _{st}$ as $\rho (r)$ and $P_{st}$ as $P_r(r).$ The
tangential pressure is 
\begin{equation}
\label{stattangential}P_{\bot }(r)-P_r(r)=\frac r2\left[ P_r\right]
_1+\left( \frac{\rho +P_r}2\right) \left( \frac{m(r)+4\pi r^3P_r}{r-2m(r)}%
\right) 
\end{equation}
$$
=\frac 3{10}\frac{kP_c}{a^2}\alpha r^4\left( 1-\frac{r^2}{a^2}\right) ^n+%
\frac{r^2}{2\left( 1-\frac{2m}r\right) }\Phi , 
$$
where%
$$
\Phi =2P_c\left( 1-2\frac mr\right) ^2\left[ 2\pi P_c\left( 1-\frac{r^2}{a^2}%
\right) ^{2n}-\frac n{a^2}\left( 1-\frac{r^2}{a^2}\right) ^{n-1}\right] 
$$
\begin{equation}
+\frac{\alpha \rho _c}2\left( 1-\frac{3k}5\frac{r^2}{a^2}\right) \left( 1-k%
\frac{r^2}{a^2}\right) , 
\end{equation}
with 
\begin{equation}
\label{alfa}\alpha =\frac{8\pi \rho _c}3. 
\end{equation}

In applying the HJR method (second point of the algorithm) we assume that
the effective variables $\widetilde{\rho }(u,r)$ and $\widetilde{P}(u,r)$
have the same $r$-dependence than the physical quantities $\rho (r)$ and $%
P(r).$ The time dependence of these ones is introduced by two arbitrary
functions of time $K(u)$ and $G(u).$ The form proposed here is 
\begin{equation}
\label{tildeenergy}\widetilde{\rho }=\widetilde{\rho }_c(u)\left( 1-K(u)%
\frac{r^2}{a^2}\right) , 
\end{equation}
and 
\begin{equation}
\label{tildepressure}\widetilde{P}=\widetilde{P}_c(u)\left( 1-2\frac{%
\widetilde{m}}r\right) \left( 1-G(u)\frac{r^2}{a^2}\right) ^n, 
\end{equation}
where 
\begin{equation}
\widetilde{\rho }_c(u)=\rho _c\frac{K(u)}{K(0)}\equiv \rho _c\frac K{K_o}, 
\end{equation}
\begin{equation}
\widetilde{P}_c(u)=P_c\frac{K(u)}{K(0)}\equiv P_c\frac K{K_o}. 
\end{equation}
The central energy density and radial pressure in the final equilibrium
state must differ from their values in the initial static case. Thus, we
introduce their {\it effective }variables $\widetilde{\rho }_c$ and $%
\widetilde{P}_c$. These variables do not enjoy any physical meaning during
the collapse, but they coincide with the central energy density and radial
pressure in the static case. Because of this, it is necessary to introduce a
time dependence in $\widetilde{\rho }_c$ and $\widetilde{P}_c.$ The
correctness of expressions (\ref{tildeenergy}) and (\ref{tildepressure}) may
be seen {\it a posteriori} at the sight of the obtained results -section V
below. As mentioned in subsection II E, the junction condition (\ref{jc5})
allows us to relate both unknown functions of time $K$ and $G$. Using (\ref
{tildeenergy}) and (\ref{tildepressure}) in (\ref{jc5}) and by means of (\ref
{adime}) they are related by the expression 
\begin{equation}
\label{G}\left( 1-G\right) ^n=\frac{\left( 1-\Omega \right) \left(
1-K\right) }{F\Omega \lambda }. 
\end{equation}

To find an expression for $\widetilde{m}(u,r)$ and $\beta (u,r)$ (third
point) we resort to the expressions (\ref{masa}) and (\ref{betatil}). In
virtue of (\ref{tildeenergy}) the mass function can be written as 
\begin{equation}
\label{masa1}\widetilde{m}=\int_0^r4\pi r^2\widetilde{\rho }dr=\frac{%
\widetilde{\alpha }r^3}2\left( 1-\frac{3K}5\frac{r^2}{A^2}\right) , 
\end{equation}
where 
\begin{equation}
\widetilde{\alpha }=\frac{8\pi \widetilde{\rho }_c}3=\alpha \frac K{K_o} 
\end{equation}
and the radius of the star $a$ is written from (\ref{adime}) as $A$ because,
without loss of generality, the initial mass is taken as unity. The
expression of $\beta (u,r),$ by means of (\ref{tildeenergy}) and (\ref
{tildepressure}), reads 
\begin{equation}
\label{beta1}\beta =\frac{3\lambda \widetilde{\alpha }A^2}{8G(n+1)}\left[
\left( 1-G\right) ^{n+1}-\left( 1-G\frac{r^2}{A^2}\right) ^{n+1}\right] -%
\frac 5{16}\ln \left[ \frac{1-\widetilde{\alpha }r^2+%
%TCIMACRO{\dfrac{3K\widetilde{\alpha }}{5A^2} }
%BeginExpansion
{\displaystyle {3K\widetilde{\alpha } \over 5A^2}}
%EndExpansion
r^4}{1-\widetilde{\alpha }A^2+%
%TCIMACRO{\dfrac{3K\widetilde{\alpha }}5 }
%BeginExpansion
{\displaystyle {3K\widetilde{\alpha } \over 5}}
%EndExpansion
A^2}\right] +\frac{\widetilde{\alpha }}{16}{\cal I}, 
\end{equation}
where ${\cal I}$ may take different values depending on the relation between 
$\widetilde{\alpha }$ and $K/A^2$:

\begin{itemize}
\item  For $%
%TCIMACRO{\dfrac{12K}{5A^2}}
%BeginExpansion
{\displaystyle {12K \over 5A^2}}
%EndExpansion
>\widetilde{\alpha }$,

${\cal I}$ takes the form 
\begin{equation}
{\cal I}=\frac 2\xi \left[ \tan ^{-1}\left( \frac{\widetilde{\alpha }\left(
6Kr^2-5A^2\right) }{5A^2\xi }\right) -\tan ^{-1}\left( \frac{\widetilde{%
\alpha }\left( 6K-5\right) }{5\xi }\right) \right] , 
\end{equation}
where 
\begin{equation}
\xi \equiv \sqrt{\frac{12\widetilde{\alpha }K}{5A^2}-\widetilde{\alpha }^2}. 
\end{equation}

\item  For $\widetilde{\alpha }>%
%TCIMACRO{\dfrac{12K}{5A^2}}
%BeginExpansion
{\displaystyle {12K \over 5A^2}}
%EndExpansion
,$%
\begin{equation}
{\cal I}=\frac 1\xi \ln \left[ \left( \frac{\widetilde{\alpha }\left(
6Kr^2-5A^2\right) -5A^2\xi }{\widetilde{\alpha }\left( 6Kr^2-5A^2\right)
+5A^2\xi }\right) \left( \frac{\widetilde{\alpha }\left( 6K-5\right) +5\xi }{%
\widetilde{\alpha }\left( 6K-5\right) -5\xi }\right) \right] , 
\end{equation}
and 
\begin{equation}
\xi \equiv \sqrt{\widetilde{\alpha }^2-\frac{12\widetilde{\alpha }K}{5A^2}}. 
\end{equation}

\item  For $\widetilde{\alpha }=%
%TCIMACRO{\dfrac{12K}{5A^2}}
%BeginExpansion
{\displaystyle {12K \over 5A^2}}
%EndExpansion
$,

${\cal I}$ takes a more simple form 
\begin{equation}
{\cal I}=-\frac{16}{\widetilde{\alpha }^3}\left[ \frac 1{\widetilde{\alpha }%
r^2-2}-\frac 1{\widetilde{\alpha }A^2-2}\right] . 
\end{equation}
\end{itemize}

Following the fourth point of the HJR method it is necessary to relate the
unknown functions of time $K$ and $G$ with the functions $A$, $\Omega $, $F$
or $L.$ To this end, we evaluate the expression (\ref{masa1}) in the surface
of the star, 
\begin{equation}
M=\widetilde{m}_a=\frac K{K_o}\left( \frac{\alpha A^3}2\right) \left( 1-%
\frac{3K}5\right) . 
\end{equation}
On the other hand in virtue of (\ref{adime}) the total mass of the system is 
\begin{equation}
M=\frac A2\left( 1-F\right) . 
\end{equation}
Equating the two last expressions it is possible to find the dependence of $%
K $ on $A$ and $F,$%
\begin{equation}
\label{K}K(u)=K(A,F)=\frac 56\left[ 1\pm \sqrt{1-\left( \frac{12K_o}{5\alpha 
}\right) \left( \frac{1-F}{A^2}\right) }\right] . 
\end{equation}
The dependence of $G$ on time (or equivalently on $A$, $F$ and $\Omega $)
can be established by substituting the last expression in (\ref{G}). Thus,
introducing the expressions $K(A,F)$ and $G(A,F,\Omega )$ in (\ref
{tildeenergy}), (\ref{tildepressure}), (\ref{masa1}) and (\ref{beta1}), $%
\widetilde{\rho }$, $\widetilde{P}$, $\widetilde{m}$ and $\beta $ are known
throughout the star and for all times as functions of $r$ and $A$, $F$ and $%
\Omega .$

To ascertain the physically meaningful sign in (\ref{K}), we evaluate it at
the initial time, 
\begin{equation}
\label{anterior}\frac 65K_o-1=\pm \sqrt{1-\frac{24K_o}{5\alpha A_o^3}}, 
\end{equation}
where we have made use of (\ref{adime}) with $\widetilde{m}(0)=1.$
Therefore, if the initial value of the function $K$ exceeds $5/6$ we must
adopt the positive sign in (\ref{K}), and the negative one otherwise.

The functions $A$, $F$ and $\Omega $ can be found from the system of surface
equations if the luminosity is known (point five of the method). The first
two surface equations (\ref{se1}) and (\ref{se2}), are model independent. In
order to find a valid expression for the third one we resort to expression
given in the last point of the algorithm (\ref{ptmpr}) which in virtue of (%
\ref{tildeenergy}), (\ref{tildepressure}) and (\ref{masa1}) can be written
as 
\begin{equation}
P_{\bot }(u,r)-P_r(u,r)=\frac 3{10}\frac{k\widetilde{P}_c}{a^2}\widetilde{%
\alpha }r^4\left( 1-G\frac{r^2}{a^2}\right) ^n+\frac{r^2}{2\left( 1-\frac{2%
\widetilde{m}}r\right) }\widetilde{\Phi }, 
\end{equation}
where%
$$
\widetilde{\Phi }=2\widetilde{P}_c\left( 1-2\frac{\widetilde{m}}r\right)
^2\left[ 2\pi \widetilde{P}_c\left( 1-G\frac{r^2}{a^2}\right) ^{2n}-\frac{nG%
}{a^2}\left( 1-G\frac{r^2}{a^2}\right) ^{n-1}\right] 
$$
\begin{equation}
+\frac{\widetilde{\alpha }\widetilde{\rho }_c}2\left( 1-\frac{3K}5\frac{r^2}{%
a^2}\right) \left( 1-K\frac{r^2}{a^2}\right) . 
\end{equation}
Therefore, the system of surface equations is, 
\begin{equation}
\label{seg1}\stackrel{\cdot }{A}=F(\Omega -1), 
\end{equation}
\begin{equation}
\label{seg2}\stackrel{\cdot }{F}=\frac 1A\left[ 2L+F(1-F)(\Omega -1)\right]
, 
\end{equation}
and 
\begin{equation}
\label{seg3}\stackrel{\cdot }{\Omega }=-\frac{\stackrel{\cdot }{F}}F\Omega +%
\frac{\stackrel{\cdot }{K}}K\frac{(1-2K)}{(1-K)}\Omega +\frac{4L\Omega ^2}{3%
\widetilde{\alpha }A^3(2\Omega -1)(1-K)}+\Omega (1-\Omega )\xi , 
\end{equation}
where 
\begin{equation}
\xi =\frac{3\widetilde{\alpha }}2A(1-K)\left( \frac{3\Omega -1}\Omega
\right) -\frac{3+F}{2A}+\frac{2F\Omega }{A(1-K)}(\Psi -K), 
\end{equation}
$$
\Psi =\frac 3{10}\lambda \widetilde{\alpha }A^2K\left( 1-G\right) ^n 
$$
\begin{equation}
+\frac{A^2}{2F}\left[ \frac{3\widetilde{\alpha }}2\lambda ^2F^2\left(
1-G\right) ^{2n}-\frac{2n\lambda G}{A^2}F^2\left( 1-G\right) ^{n-1}+\frac{%
\widetilde{\alpha }}2\left( 1-\frac{3K}5\right) \left( 1-K\right) \right] . 
\end{equation}
This system of differential equations can be solved for a given set of
initial values of $A$, $F$ and $\Omega $ imposing a boundary condition (in
this case the luminosity $L$).

The initial central energy density can be written, by means of (\ref{alfa})
and (\ref{anterior}), as a function of initial radius and $K_o,$%
\begin{equation}
\label{centralenergy}\rho _c=\frac{15}{4\pi A_o^3\left( 5-3K_o\right) }. 
\end{equation}
Thus, a large radius implies a low central energy density. Moreover, the
positive character of the discriminant in (\ref{K}) implies some
restrictions about the evolution of the total mass $M$ of the star. During
the collapse it must fulfill the inequality 
\begin{equation}
M\leq \frac{5\alpha }{24K_o}A^3. 
\end{equation}

In the static case a non negative value of $K$ ({\it i.e.} $K_o$), ensures
the condition $d\rho /dr\leq 0$, while condition $K_o\leq 1$ leads to a
positive energy density. The effective energy density lacks of physical
meaning when the star is collapsing. Therefore, the last argument cannot be
applied to decide the range of values of function $K$ here. Evaluating (\ref
{rotilde}) in the surface and using (\ref{jc6}) and (\ref{adime}) we obtain 
\begin{equation}
\widetilde{\rho }_a=\frac \Omega {2\Omega -1}\left( \rho _a-P_{ra}\right) . 
\end{equation}
$\Omega $ lies in the range $1/2<\Omega <\infty $, further the condition $%
\rho \geq P_r$ must be fulfilled. Thus, from the last expression, $%
\widetilde{\rho }_a\geq 0.$ Applying this restriction to (\ref{tildeenergy})
evaluated in the surface we obtain 
\begin{equation}
K\left( 1-K\right) \geq 0\Longrightarrow 0\leq K\leq 1, 
\end{equation}
therefore the range of values physically admissible for $K$ coincides with
the static one.

A last remark is in order before numerically solving the system of surface
equations. Note that the generalization adopted for the tangential pressure (%
\ref{ptmpr}) does not allow explosive models \cite{CoHeEsWi82}.

\subsection{Evolution of the GM configuration}

We now are prepared to study the evolution of the initial static fluid
distribution described above in a particular case. To this end, it is
necessary to adopt a specific initial configuration. The neutrinos can be
studied in the diffusive regime if the density is higher than $10^{11}\sim
10^{12}$ g cm$^{-3}$. As mentioned in the introduction, the neutrino
trapping is an important source of viscosity. Because of that, in our
diffusive model, the energy density in the surface must be, at least, about $%
10^{12}$ g cm$^{-3}.$ On the other hand a neutron star with a radius of
about $10$ km, has a central density not far from $10^{15}$ g cm$^{-3}$ \cite
{ShTe83,Demianski85}. In this diffusive model, the energy density in the
center is a thousand times the corresponding to the surface. Thus, from (\ref
{statenergy}), $K_o=0.999.$ From expression (\ref{centralenergy}) one can
find the initial radius for a given initial stellar mass. Introducing the
usual dimensions in (\ref{centralenergy}) we obtain 
\begin{equation}
\label{centraltrue}\rho _c=\frac{c^2}G\frac{15}{4\pi A_o^3\left(
5-3K_o\right) M_o^2},
\end{equation}
where the initial mass $M_o$ is given in geometrized units. Assuming $A_o=6$%
, the initial mass is close to $1.3$ M$_{\odot }.$ The corresponding initial
radius can be found from (\ref{adime}) 
\begin{equation}
a_o=A_oM_o\simeq 11521\text{ m.}
\end{equation}
This values for the initial radius and mass are in the range of the usually
accepted as typical for neutron stars. So we specifically adopt as initial
configuration%
$$
\rho _c\simeq 1.01\times 10^{15}\text{ g cm}^{-3}, 
$$
$$
\rho _a\simeq 1.01\times 10^{12}\text{ g cm}^{-3}, 
$$
$$
M_o=1.3\text{ M}_{\odot } 
$$
and 
\begin{equation}
a_o\simeq 11521\text{ m,}
\end{equation}
corresponding to a dense neutron star in which the diffusion approximation
for neutrinos holds.

To completely determine the initial conditions it is necessary to give a
relationship between the energy density and pressure in the center of the
star (\ref{pcversusroc}). Also the parameter $n$ occurring in the expression
for pressure in the static case (\ref{statpressure}) must be imposed. If we
assume the center of the star as a highly relativistic Fermi gas, then $%
\lambda =1/3$. In addition we take $n=1$ for simplicity.

Therefore, the initial conditions for the system of surface equations (\ref
{seg1}-\ref{seg3}) are 
\begin{equation}
\label{coninidifu}A_o=6,\ F_o=\frac 23,\ \Omega _o=1,\ K_o=0.999,\ P_c=\frac 
13\rho _c\ \text{and }\ n=1. 
\end{equation}
These ones correspond to a neutron star, initially at rest, with a redshift (%
$z_a=1/\sqrt{F}-1$) close to $0.22.$

The boundary condition, necessary to solve the system of surface equations,
is supplied by the luminosity. We assume a Gaussian pulse for $L$ centered
in $u=u_{peak}$, and width $\Lambda ,$%
\begin{equation}
L=-\stackrel{\cdot }{M}=\frac{M_r}{\Lambda \sqrt{2\pi }}\exp \left( -\frac 12%
\left[ \frac{u-u_{peak}}\Lambda \right] ^2\right) . 
\end{equation}
Note that $M_r$ is the total energy radiated by the star in the collapse.
The values adopted for $M_r$, $\Lambda $ and $u_{peak}$ are $0.01M_o$, $15$
and $150$ respectively. We have imposed the energy conditions for imperfect
fluids \cite{KoSaTs88} and the restriction $-1<\omega <1$.

It is worth to emphasize the behavior of this model close to surface of the
star. The surface collapses more slowly than the adjacent layers (figure1).
It may be traced to the profile of the pressure gradient in this region. As
seen in figure 2, its absolute value is larger in the surface than in the
adjacent layers, which find a lesser resistance to collapse. Similar
behavior is observed for lower values of $K_o$. Nevertheless, for low $K_o$
the variation in the pressure gradient is smaller than for large $K_o$ and,
consequently, the difference between the velocity of surface and adjacent
layers is not too high.

The value of $K_o$ largely determines the radius of the star once the
collapse is finished. The higher $K_o$, the lower final radius. This
behavior can be related to the fact that values of $K_o$ far from unity
follows from an equation of state close to incompressibility. Thus, in these
cases, the velocity of the surface is not very high and possible variations
in the radius are restricted. The highest variations in energy density and
radial pressure take place in the most external layers, which comprise
approximately a thousand meters below the surface (figures 3 and 4).

The evolution of the energy density in the surface is depicted in figure 5.
For $K_o=0.9$ ($\rho _a(u=0)\simeq 8.78\times 10^{13}$ g cm$^{-3}$) it
varies about $6.5\%$ , while for $K_o=0.999$ the final energy density in the
surface gets nine times its initial value. This behavior, with that of the
velocity, hints that the compression of the star is higher in the shells
near the surface. This is likely due to the radiation generated in the
collapse. As it is showed in figure 6 the heat flow is more intense in the
inner shells than in the more external ones.

Initially the viscous pressure vanishes due to the absence of radiation.
This condition is also fulfilled in the equilibrium situation once the
collapse is over. Its contribution to the tangential pressure during the
collapse must be established by means of the transport equations.

\subsection{Boundary and initial conditions for the transport equations}

We have solved the transport equation for the heat flow (\ref{difeflujo})
for different boundary conditions. The interior solution found for the
temperature ceases to be sensitive to the boundary condition about $500\sim
700$ m below the surface. This hints that the energy of the emergent
neutrinos is not correlated with its energy beyond this external region and,
consequently, most of the neutrinos escaping the star have been generated in
the inner limit of this region. This agrees with the idea of neutrinosphere 
\cite{HaKa94}. A possible cause of this behavior can be found in the abrupt
decrease of the heat flow close to surface, that follows from the similar
behavior shown by the energy density and radial pressure in this zone. As
the model of Gokhroo and Mehra, the BBS model and others based on nuclear
physics show the same features \cite{ShTe83,Demianski85}. Thus, the
neutrinosphere seems to occur whenever the adopted model is not highly
idealized.

Aimed to establish the boundary condition for the temperature we introduce
the effective temperature,{\it \ $T_{eff}$. }This one is usually defined by
means of 
\begin{equation}
\label{Eteff}E=\left[ 4\pi r^2\sigma T_{eff}^4\right] _{r=a}, 
\end{equation}
where $\sigma =bc/4,$ and 
\begin{equation}
\label{Enocom}E=\left[ 4\pi r^2\varepsilon \right] _{r=a} 
\end{equation}
is the luminosity as measured by the non-comoving observer momentarily
located on the surface. It is to say, {\it $T_{eff}$} would be the
temperature in the surface of the star if it would radiate as a black body.
The idea of effective temperature -see for instance \cite[page 586]{ShTe83}
and \cite[page 295]{HaKa94}- is applicable to the most external layers of
the star. It is related with the material temperature by the expression

\begin{equation}
\label{t4tau}T^4=\frac 12T_{eff}^4\left( 1+\frac 32\tau \right) , 
\end{equation}
where $\tau $ is the optical depth ({\it i.e. }$d\tau =-dr/\lambda _{eff}$).

According to last expression, if $\tau =2/3$ the effective temperature
coincides with the material one$.$ Thus, most of the emergent neutrinos are
generated in a shell close to $\tau \sim 2/3.$ This one is, in the model
under consideration, about $500$ meters below the surface. At the surface $%
\tau $ vanishes. Hence, the material temperature of the surface can be
written as 
\begin{equation}
\label{tasupert}\left[ T^4\right] _a=\frac 12\left[ T_{eff}^4\right] _a.
\end{equation}

The boundary condition can be found from last expression together with (\ref
{ehat}), (\ref{esinhat}), (\ref{Eteff}) and (\ref{tasupert}). It reads 
\begin{equation}
\label{boundaryt}T_a^4={\cal Q}_a\frac{2\Omega -1}{2\sigma }.
\end{equation}
This condition on the surface affects the evolution of the temperature in
the layers close to the surface but not the evolution of the inner
temperature (figure 7).

To find explicitly the bulk and shear viscous pressure it is necessary to
impose initial and boundary conditions. Something about the initial
condition for the viscous pressure has been noted at the end of the last
subsection. We assume that viscous processes appear because of the
interaction between matter and neutrinos. Therefore, initially $\Pi $ and $%
\pi $ must vanish because of the absence of radiation.

At the center of the star the isotropy condition ($P_{\bot }=P_r$) is
fulfilled. Thus,\ in virtue of (\ref{ptan}) the relation 
\begin{equation}
\label{prara}\lim \limits_{r\rightarrow 0}\left( \wp -\frac 32\pi \right) =0
\end{equation}
must be satisfied. And as a consequence of the regularity condition $\left[ 
\widetilde{m}\right] _{r=0}=0$, at the center of the star the generation of
neutrinos is forbidden. Therefore we assume that both viscous pressures,
vanish at $r=0$. Thus, the boundary conditions imposed on (\ref{bdifepvv})
and (\ref{bdifepvt}) are 
\begin{equation}
\left[ \Pi \right] _{r=0}=\left[ \pi \right] _{r=0}=0\qquad \forall u,
\end{equation}
and from (\ref{prara}) 
\begin{equation}
\left[ \wp \right] _{r=0}=0\quad \ \ \forall u.
\end{equation}

\subsection{Temperature and viscous pressure: main results}

We have used the Maxwell-Cattaneo transport equations instead of Eckart's
since the latter ones violate relativistic causality. The temperature in the
inner layers found by means of (\ref{difeflujo}) is the same as the one
obtained using (\ref{eck}) (i.e., Fourier's law). Nevertheless, in the
neutrinosphere the differences between both solutions cannot be neglected,
especially in the first and late stages of the collapse (figure 8). The main
difference between Maxwell-Cattaneo and Eckart heat transport equations is
the presence of the term $\tau \stackrel{\cdot }{{\cal Q}^\nu }h_\nu^\mu$, 
which introduces in (\ref{difeflujo}), the term 
\begin{equation}
f(u,r)\left( \frac{aM_o^2}\phi \right) ^{3/5}, 
\end{equation}
while ${\cal Q}^\mu $ leads to 
\begin{equation}
\frac{g(u,r)}{{\cal B}}. 
\end{equation}
In the neutrinosphere and in the case under consideration one has, $%
f(u,r)\sim 0.1$ $g(u,r)$ and ${\cal B}\sim 10^{18},$ {\it i.e.} both terms
are of about the same order when temperature approaches $10^{11}$ K. Thus,
the effects introduced by the presence of a relaxation term in heat
transport equation are important there.

The high value reached by the temperature (figure 7) validates the adopted
initial and final conditions: the approximation that in the static case the
temperature for the neutron star vanishes. Applying this condition with (\ref
{boundaryt}) to expression (\ref{difeflujo}) allows us to find the evolution
of the temperature. Once the temperature is known, the transport
coefficients can be calculated with the help of (\ref{chi}-\ref{gi}).

The bulk viscous coefficient, $\zeta $, vanishes at the center of the star
because of the state equation (\ref{coninidifu}) adopted there. In the
shells close to the center the state equation is near to $P_r=\rho /3$.
Thus, the value of $\zeta $ is less than corresponding to shear viscous
coefficient ($\eta $) (figures 9 and 10). Nevertheless, in virtue of (\ref
{bulkcoeff}), in stars with a central equation of state different from the
adopted in this case -{\it i.e.} not highly relativistic-, $\zeta $ can be
of the order of $\eta .$ On the other hand a comparison among the terms in
the evolution equations (\ref{bdifepvv}) and (\ref{bdifepvt}) points out
that the term 
\begin{equation}
\left[ \frac 13-\left( \frac{\partial P}{\partial \rho }\right) _n\right] ^2 
\end{equation}
is responsible for the main differences between the bulk and shear viscous
pressures.

In the region in which the energy density is roughly $10^{14}$ g cm$^{-3}$
we find for the shear viscosity coefficient values about $10^{28}$ dyne cm$%
^{-2}$ s$^{-1}$, which is $10^9$ times higher than the corresponding to the
interactions among electrons, protons and neutrons at the same density \cite
{FlIt79}. This underlines the importance of the neutrino trapping as a
source of viscosity in the stellar collapse. Similarly the thermal
conductivity coefficient is greater for neutrino scattering than the
mentioned \cite{FlIt79,FlIt76}.

The evolution of $\pi $ is shown in figure 11 for different layers of the
star. The shear viscosity value rises swiftly form zero, at the center of
the star, and becomes an important source of anisotropy in the core of the
star (figure 12). For distances from the center larger than $2$ km, $\pi $
can be neglected as compared with $\wp $ as a source of anisotropy.

The bulk viscous pressure is depicted in figure 13. In this model $\Pi $ can
be neglected against $\pi $ in the innermost shells. However, in the
peripheral layers $\Pi $ and $\pi $, are of the same order.

\section{Discussion}

In this paper we have used the HJR method to generate non static solution
departing from the static anisotropic fluid distribution of Gokhroo and
Mehra \cite{GoMe94}. This model shows an initial non uniform energy density.
The presence in the expression for the energy density of the parameter $K$
allows us to model without difficulties the initial features of the star
according to the radiation limit adopted on it (diffusion or free-streaming
out).

During the collapse huge quantities of neutrinos are generated. These ones
transport thermal energy from the most interior regions to the exterior
layers \cite{ShTe83,Bruenn85,Bahcall89}. Because of the high densities and
temperature, the neutrinos interact with matter. In their trip toward the
surface they get thermalized and drive the star to a new equilibrium state.
The most important thermalization processes are the absorption of neutrinos
by neutrons and collisions between neutrinos and electrons. Consequently we
have chosen a density range according to these interactions ($\rho _c\sim
10^{15}$ g cm$^{-3},$ $\rho _a\sim 10^{12}$ g cm$^{-3}$), allowing us to
treat the radiation in the diffusive limit.

The density and pressure profile in the central layers vary little during
the collapse. Therefore, the behavior of the zones nearby the surface is
perhaps the most important. Due to the decrease, in these layers, of the
pressure gradient with respect to the surface, the velocity of collapse of
the former is larger than the corresponding to this one. This behavior also
occurs in models with a density at the surface of about $10^{14}$g cm$^{-3}$%
, though then the effect is not so intense. Also, the final radius of the
star depends on the density at the surface. The higher initial energy
density at the surface, the larger final radius. As previously mentioned,
the largest variations in energy density occur at the surface. This effect
is probably due to the importance of the radiation in the evolution of the
neutrinosphere. In that zone, the heat flow is much weaker than in the inner
regions. Nevertheless, it is possible that the radiation energy density in
it comprises a large part of total energy density.

The knowledge of the temperature of the star is the key to establish which
processes can take place in its bosom. At the variance with the heat flow,
this quantity cannot be obtained from Einstein's equations. It is absolutely
necessary then to appeal to some thermodynamic theory of irreversible
processes. We have taken a further step with respect to other analysis by
introducing the EIT in its covariant formulation \cite{Israel76,PaJoCa82},
avoiding in this way the unphysical behavior of the conventional theory \cite
{Eckart40}. The rigorous study of transport equations derived from EIT is
technically complex. Therefore, we have appealed to the so-called
''truncated'' form that reduces to the Maxwell-Cattaneo equations \cite
{JoCaLe93}. To solve the corresponding equation for temperature we have
found the mean free time of the neutrinos. In so-doing we have considered
both reactions mentioned above, if the neutrinos are generated by thermal
emission with energy near $k_BT$. In this way it is possible to establish
the dependence of the thermal conductivity coefficient on temperature ($\chi
\propto T^{3/2}$). Then, solving the equation for the evolution of the heat
flow, we found the temperature. Also we have solved the classic transport
equation, finding out considerable differences in the neutrinosphere between
both theories. These temperatures disagree from one another in the first and
last stages of the collapse.

It is worth to emphasize a peculiarity of the temperature in the innermost
layers. This one is insensitive in that zone to the boundary condition
imposed on the surface. Therefore, those neutrinos that manage to escape
from the star lack information about the temperature of the central regions,
in other words their energy in the surface is not correlated with that of
the interior. This suggests that these have been created in a region close
to the surface, the so-called neutrinosphere -\cite[page 586]{ShTe83}, 
\cite[page 295]{HaKa94}, \cite[page 17]{Chiu68}. The neutrinosphere is
commonly associated with the effective temperature, $T_{eff}$, which
coincides with the material temperature in its lower limit. Through the
introduction of $T_{eff}$ we find out the temperature at the surface. Thus,
we have obtained a credible boundary condition for the heat transport
equation. We have supposed, therefore, that the layers of the star that are
sensitive to a change in the boundary condition form the neutrinosphere with
a thickness of $500\sim 700$ meters. This length approximately coincides
with the neutrino mean free path in that zone. This fact reinforces the
hypothesis that the emergent neutrinos have been created in the interior
region that delimits the neutrinosphere.

Once the temperature was found, we went on by solving the transport
equations of the remaining dissipative flows (bulk and shear viscous
pressure). Using Weinberg expressions \cite{Weinberg71} for the viscosity
coefficients and expression (\ref{time}), we have obtained $\zeta ,\eta
\propto T^{5/2}$. This yields a value about $10^{29}$ poise for shear
viscosity coefficient. This value is much larger than the corresponding for
interactions among electrons, protons and neutrons. Although $\zeta $ can be
of the same order that $\eta $, the value we have obtained for $\zeta $ is
about one hundred times lower than $\eta $. This is because we have assumed
the center of the star as an ultrarelativistic Fermi gas and it implies an
equation of state of the form $P=\rho /3$. It is well known that for a fluid
governed by this equation the bulk viscosity coefficient vanishes \cite
{Weinberg71}. Though throughout the star this limit is not strictly
fulfilled, the relationship between the radial pressure and energy density
restricts the value of $\zeta $. In spite of this, from seven kilometers of
the center upward the bulk viscous pressure is comparable to shear one. In
an anisotropic model, shear viscous pressure is restricted by the difference
between the tangential and radial pressures. Therefore, one can expect that
its contribution to the total pressure is small. It is worth noting the
contribution of shear viscous pressure to the anisotropy. According to the
model studied here, it seems that the viscosity is responsible for an
important part of the inner anisotropy. Then, the shear viscous pressure
greatly contributes to total anisotropy in approximately the two nearest
kilometers to the center of the star. It is certain that in the most
internal zone of the star the anisotropy is less important that in the
peripheral region. Though, under certain circumstances (larger collapse
speeds) the importance of viscous pressure in connection to the anisotropy
will be extended to more afar zones.

\section*{Acknowledgments}

I would like to thank the staff of the Laboratorio de F\'\i sica Te\'orica
de la Universidad de Los Andes (M\'erida, Venezuela) for their hospitality,
especially to Lu\'\i s N\'u\~nez to notice me about reference \cite{GoMe94}
and for useful discussions at different stages of this work. Also Diego
Pav\'on is acknowledged for his encouragement and advice. This work has been
partially supported by the Programa de Formaci\'on Cient\'\i fico de la
Universidad de Los Andes (M\'erida, Venezuela) and by the Spanish Ministry
of Education under grant PB94-0718.

\newpage

\section*{Figure captions}

\begin{description}
\item[Figure 1.]  Velocity profile a for given constant Schwarzschild time ($%
T=0.9$ ms) and two different values of $K_o.$ All the figures made at
constant Schwarzschild time were constructed from expression (\ref{tiempsch}%
).

\item[Figure 2.]  Pressure gradient profile. The decrease of $\left|
P_{r1}\right| $ close to surface is, probably, the cause of the behavior
shown by the velocity in this region. Schwarzschild time $T=0.9$ ms.

\item[Figure 3.]  Relative variation of the energy density as a function of
the retarded time $u$. The corresponding value for $K_o$ is $0.999$. $\rho _o
$ denotes the initial energy density.

\item[Figure 4.]  Relative variation of the radial pressure as a function of
the retarded time $u$. $K_o=0.999$.

\item[Figure 5.]  Energy density evolution at the surface for two different
models ($K_o=0.999$ and $K_o=0.9$). Note the different scale used in each
case.

\item[Figure 6.]  Radial heat flow ${\cal Q}$ as measured by the Minkowskian
observer locally comoving with matter. $K_o=0.999$ and $T=0.95$ ms.

\item[Figure 7.]  Temperature as measured by an observed momentanously
located at the surface. Its maximum value at the surface is about $%
5.44\times 10^{11}$ K. Here, and in the following figures, ${\cal Y}_e=0.2$.

\item[Figure 8.]  Difference in the neutrinosphere between the temperature
found by means of the Eckart and Maxwell-Cattaneo transport equations.

\item[Figure 9.]  Evolution of the bulk viscosity coefficient $\zeta $.

\item[Figure 10.]  Evolution of the shear viscous coefficient $\eta .$

\item[Figure 11.]  Evolution of the shear viscous pressure in different
layers of the star.

\item[Figure 12.]  Comparative plot of the different contributions to the
anisotropy (\ref{ptan}) at $r=0.1A_o$. Lines marked A and B correspond to
terms $-3\pi /2$ and $\wp $, respectively.

\item[Figure 13.]  Evolution of the bulk viscous pressure within the star.
\end{description}

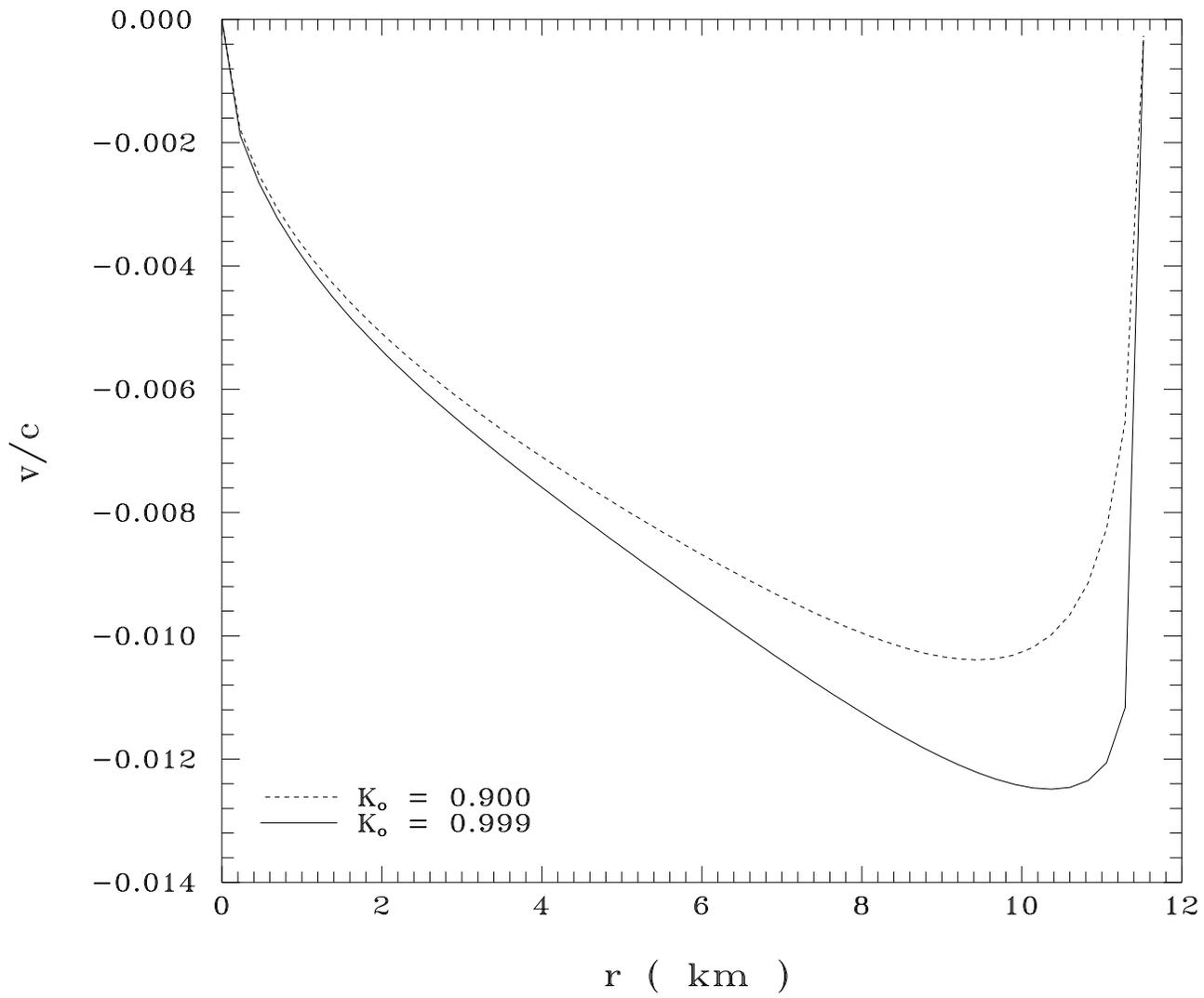
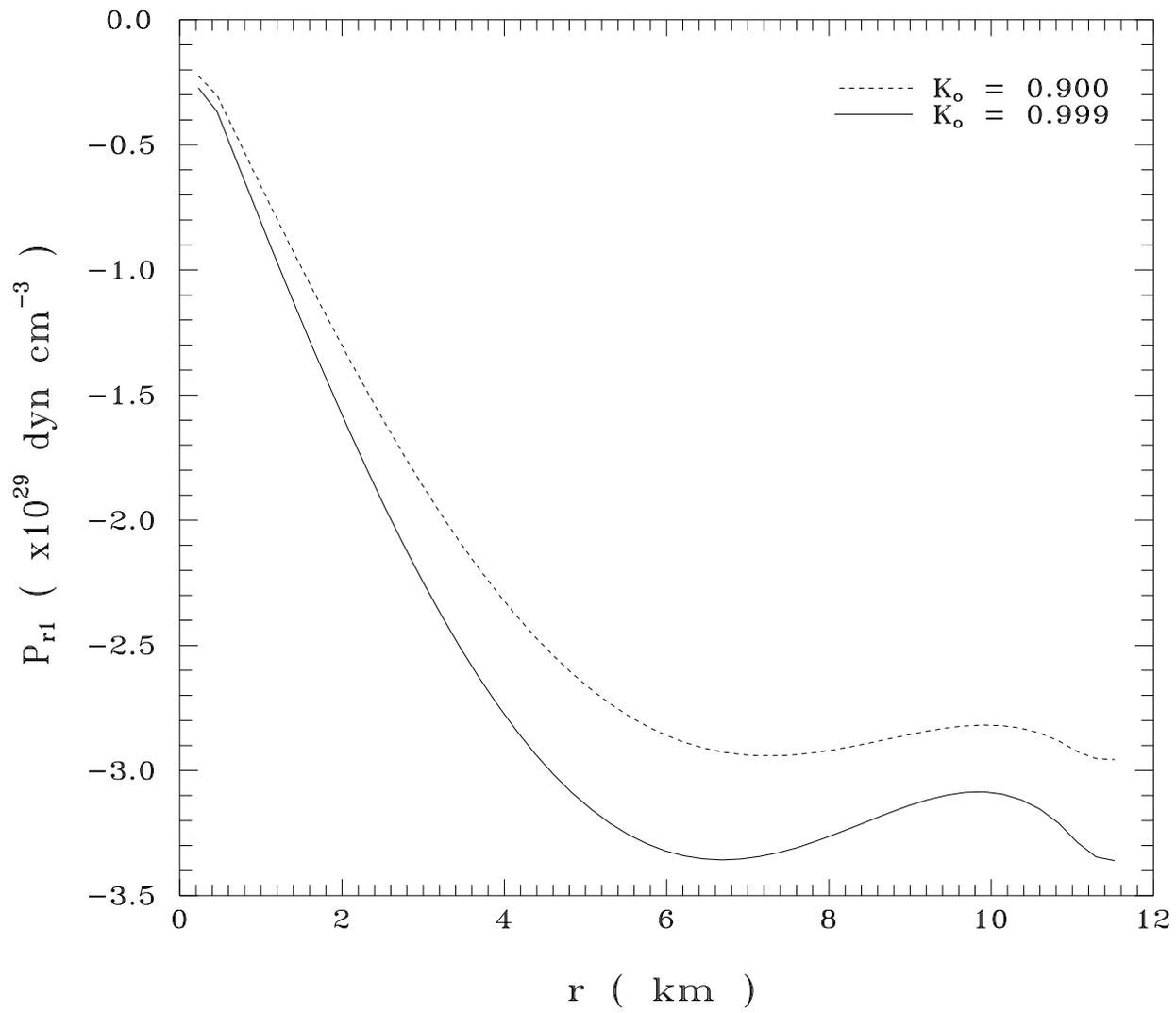
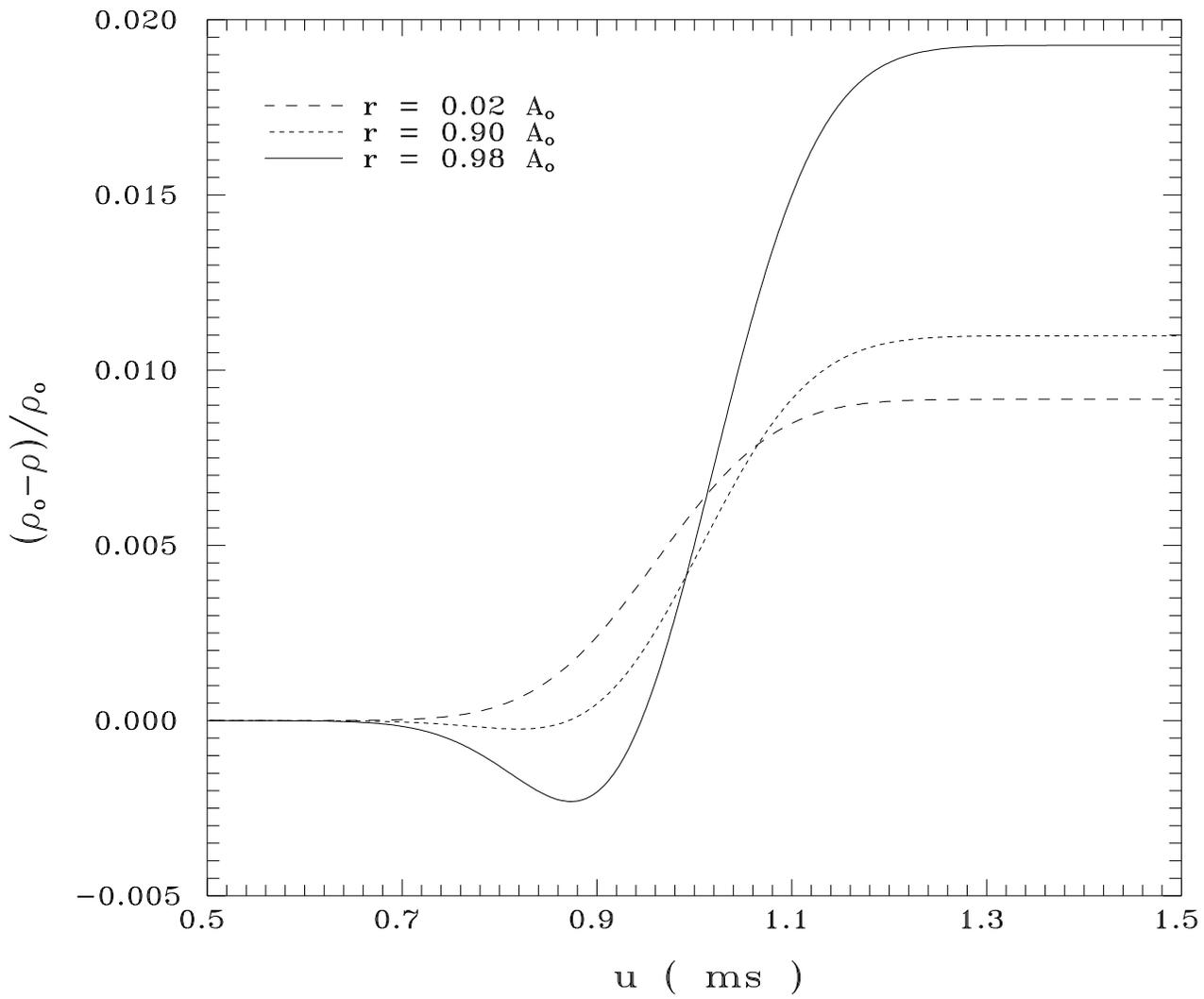
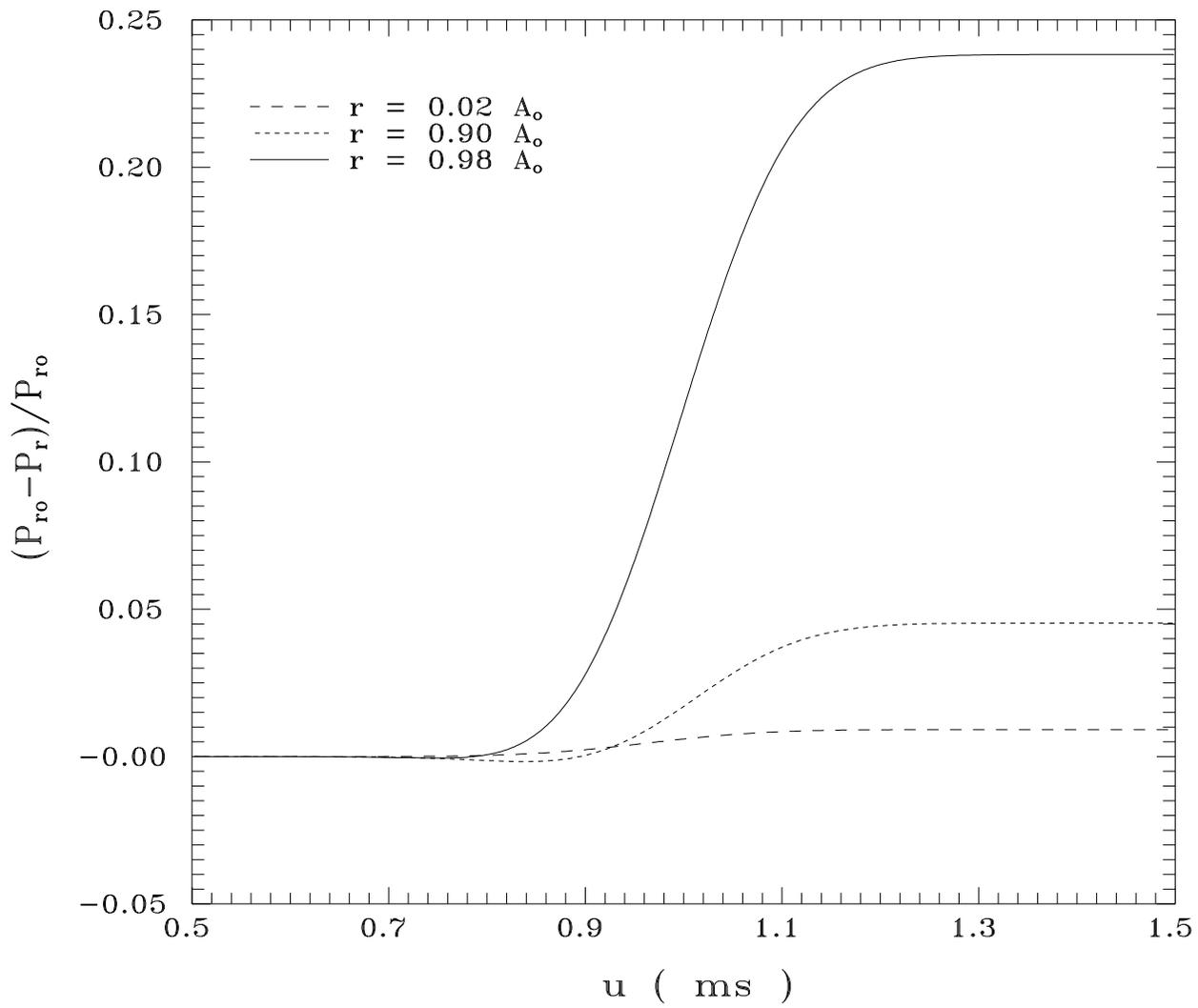
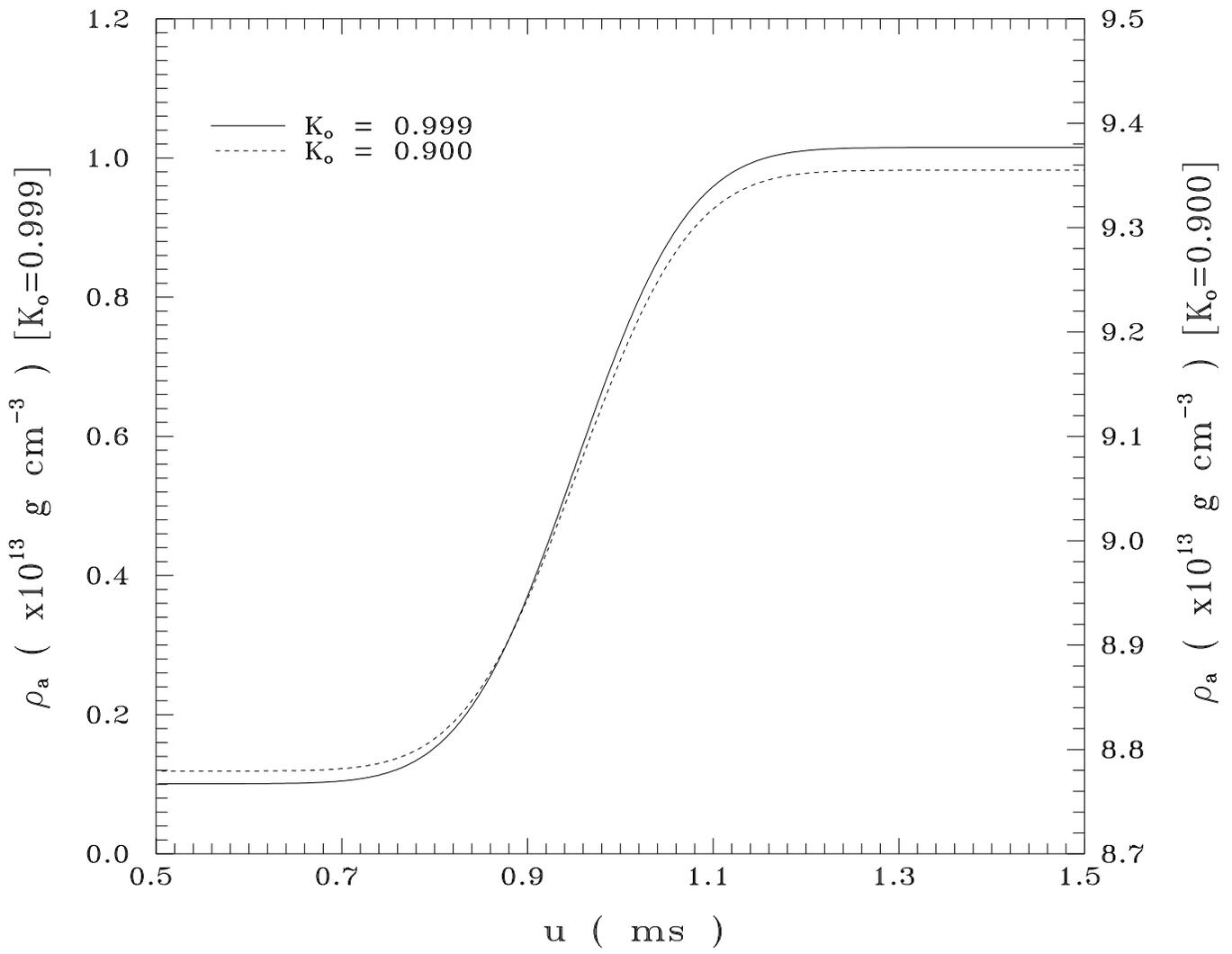
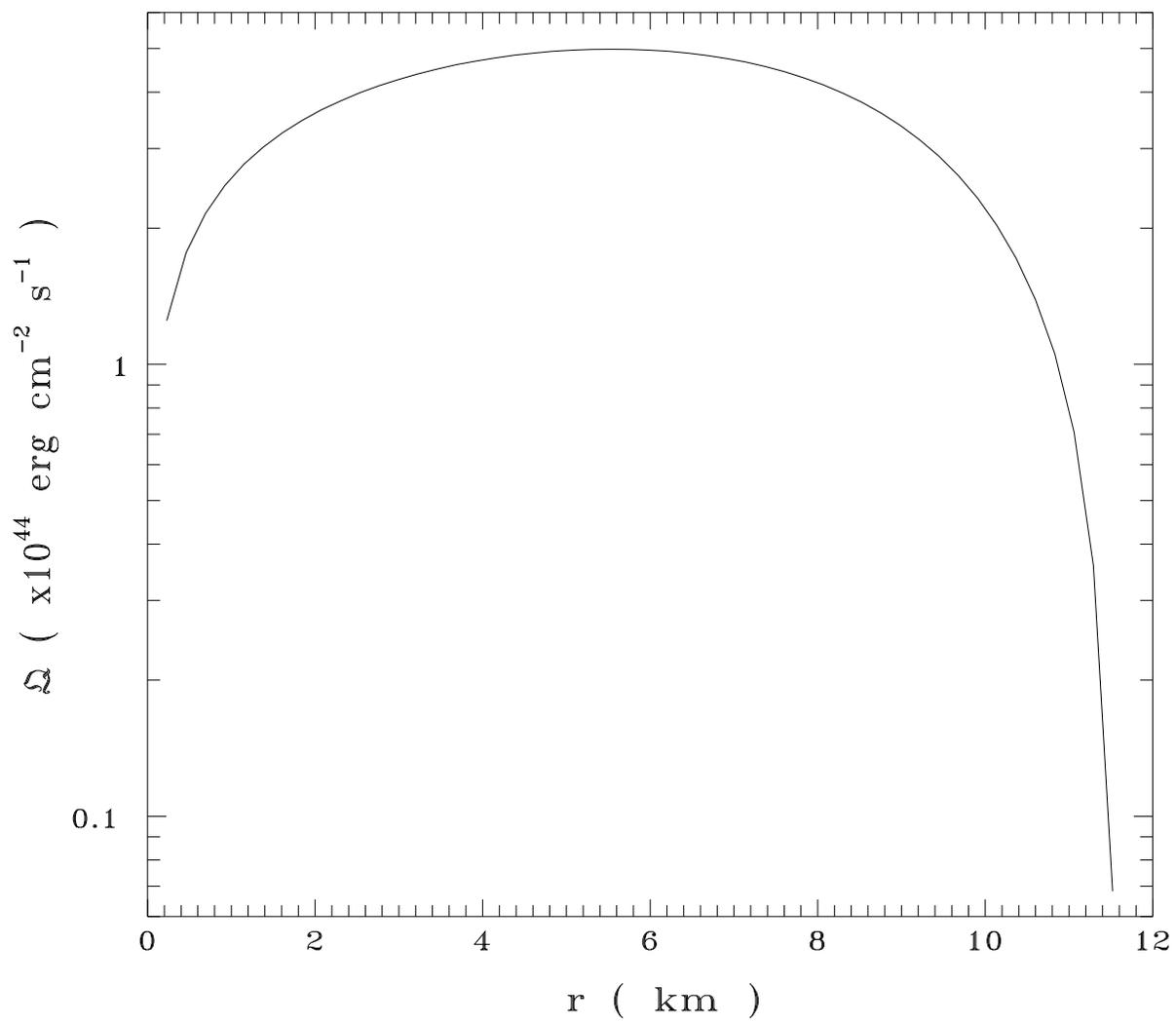
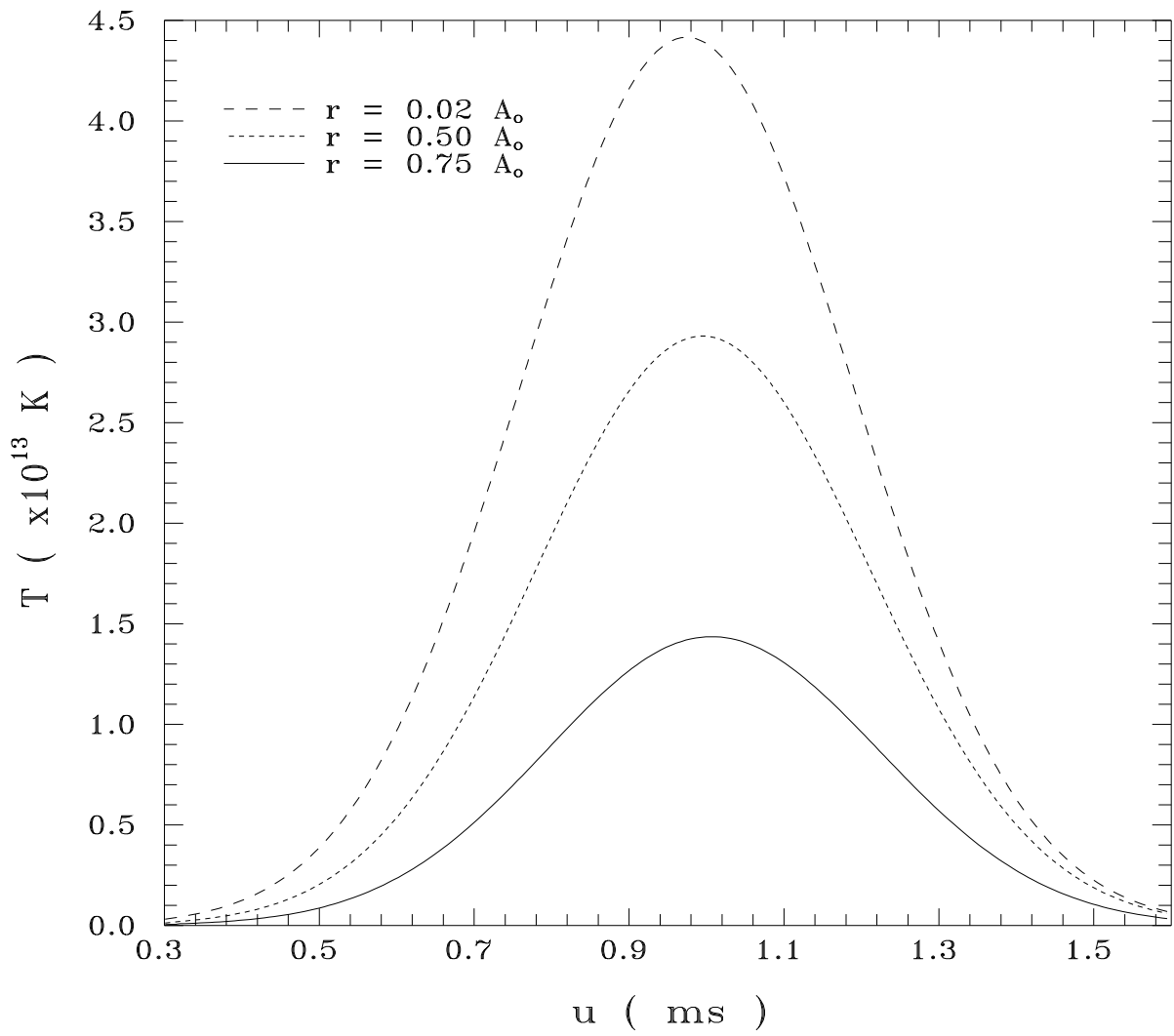
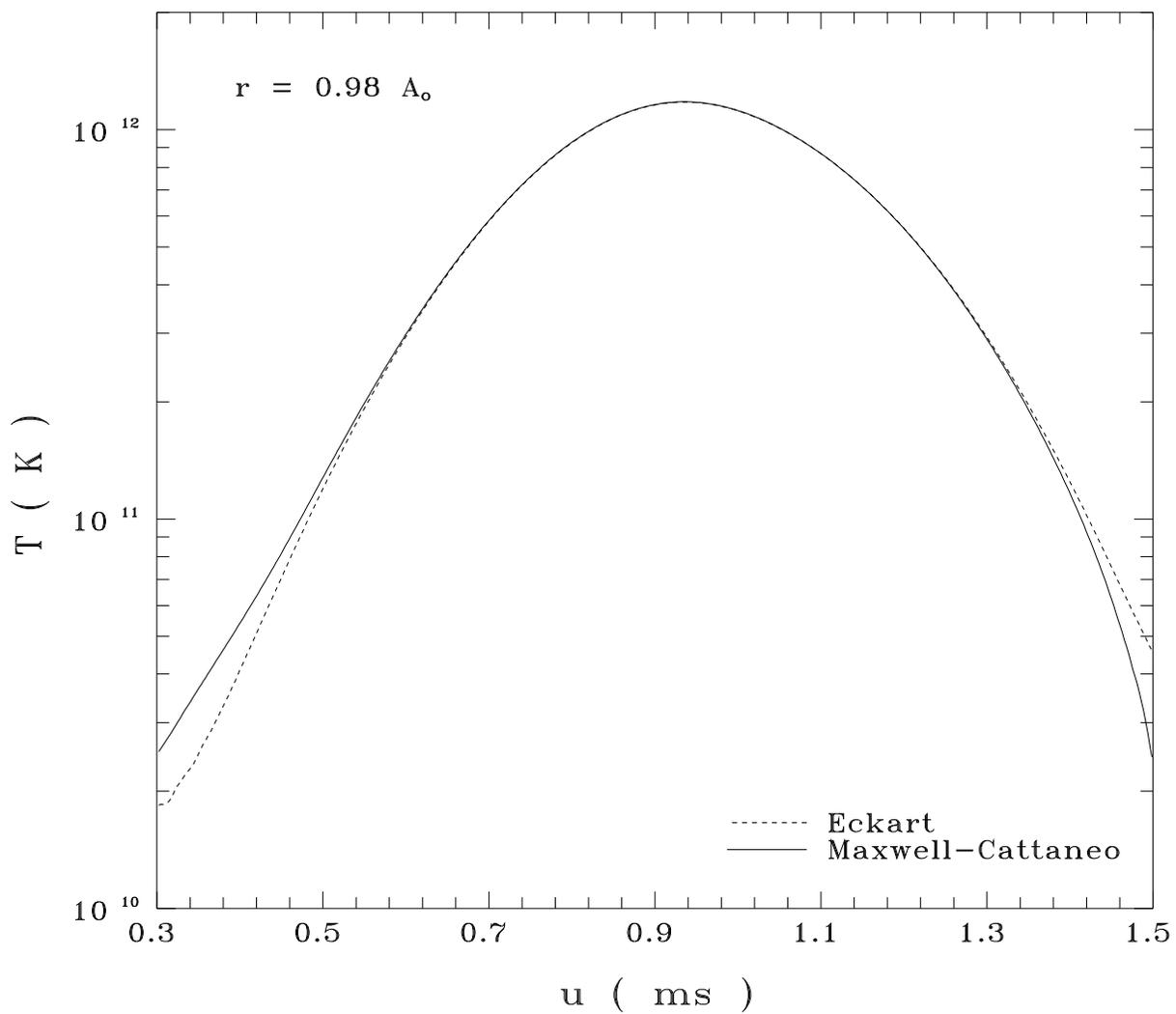
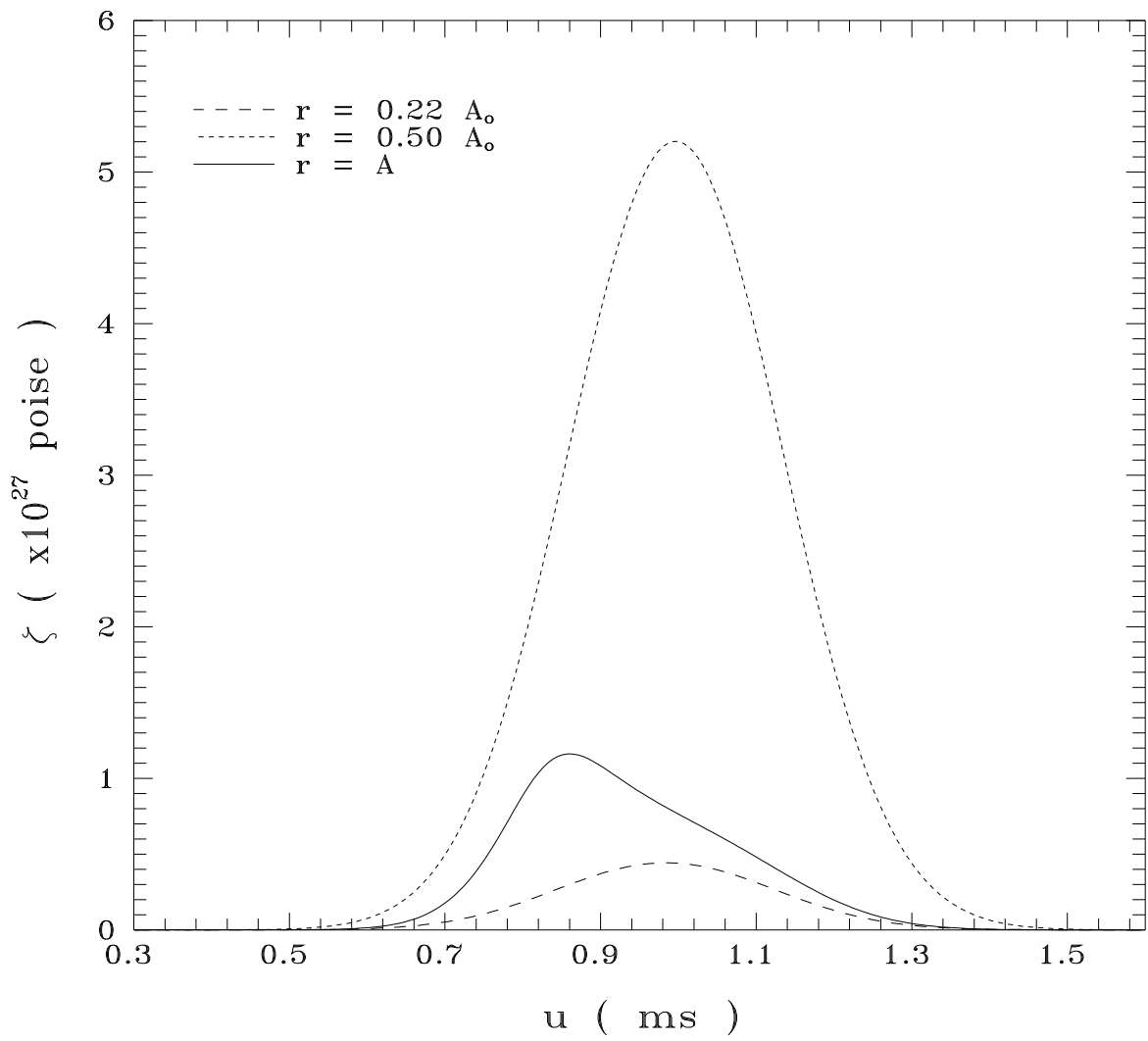
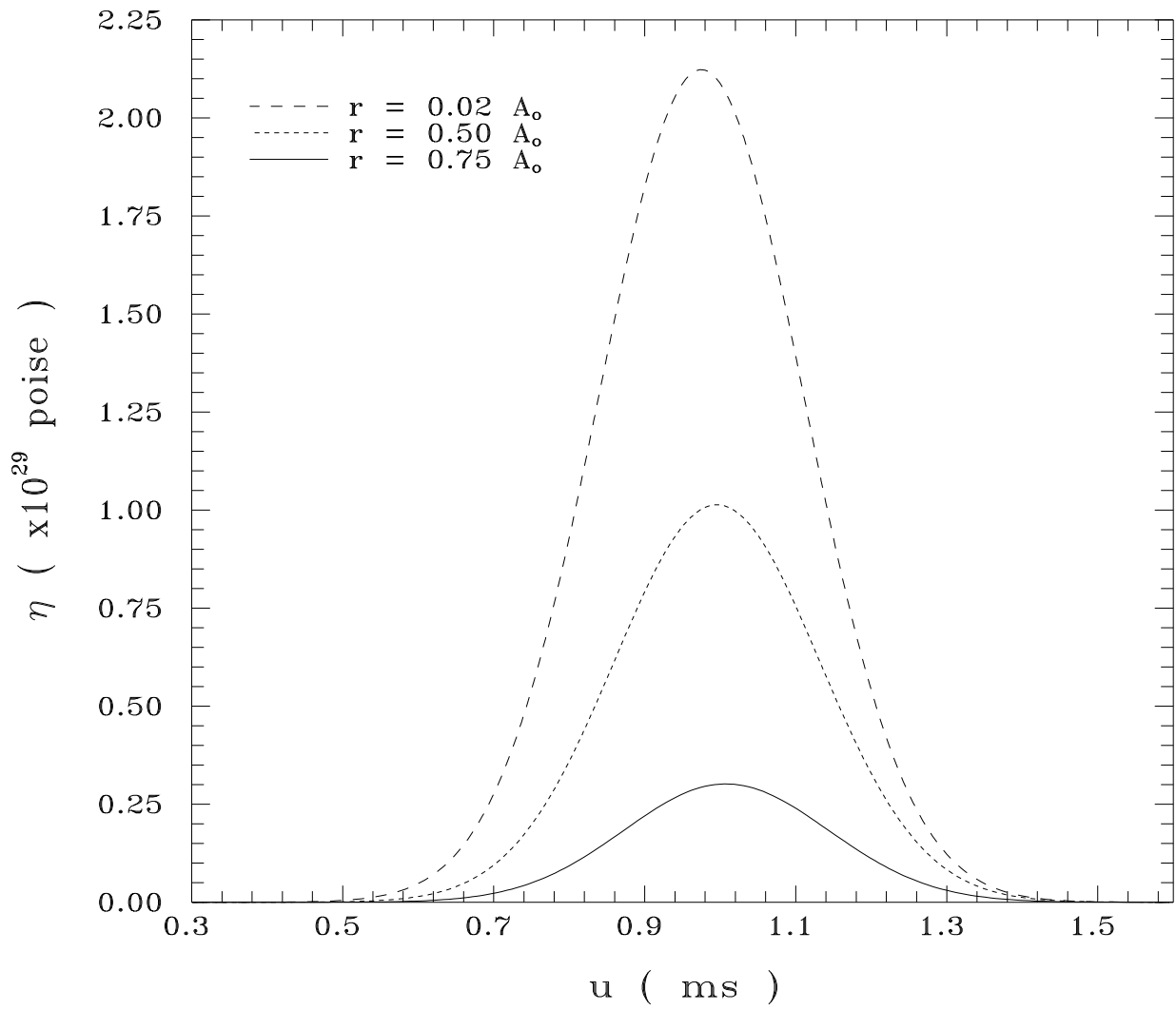
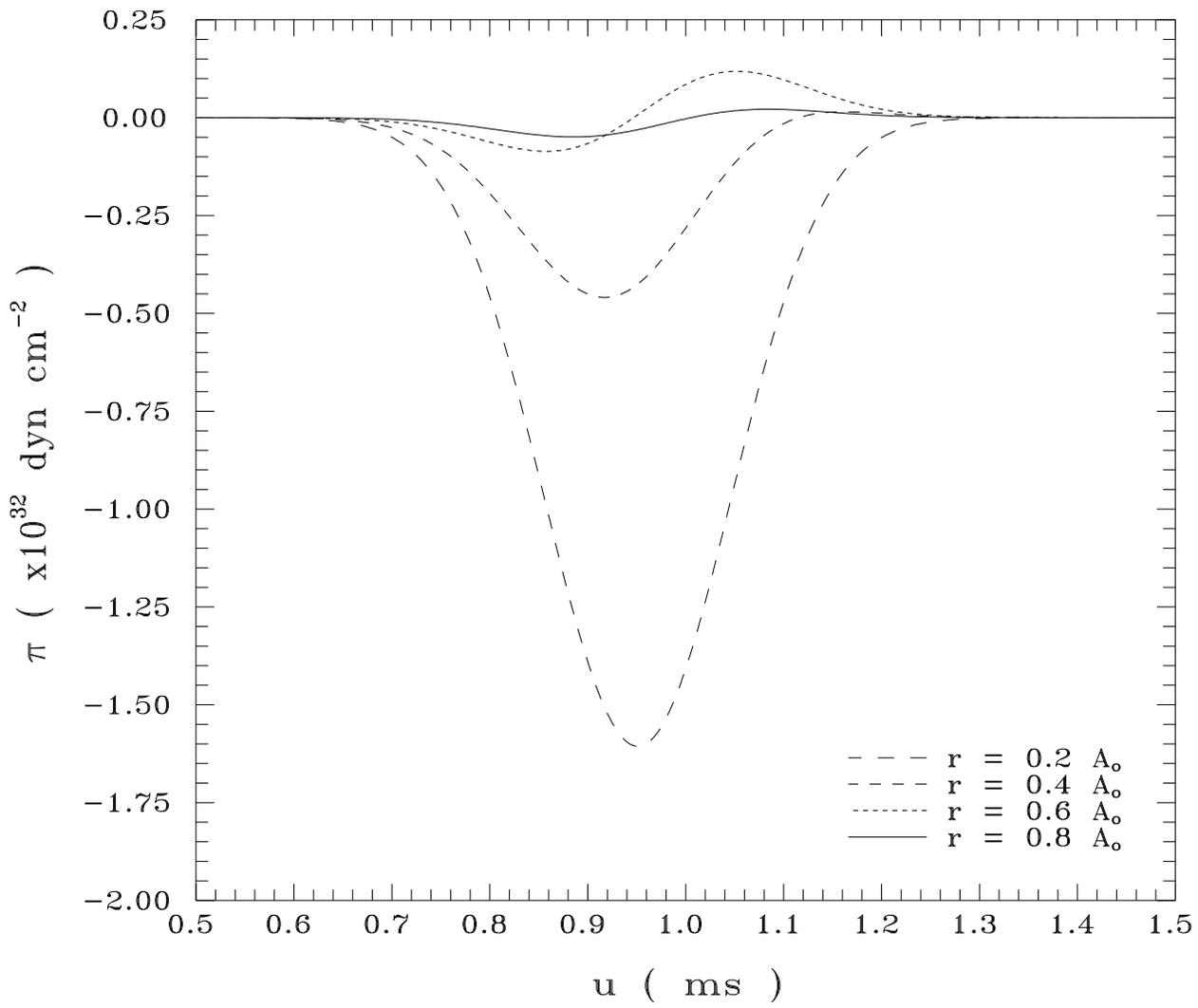
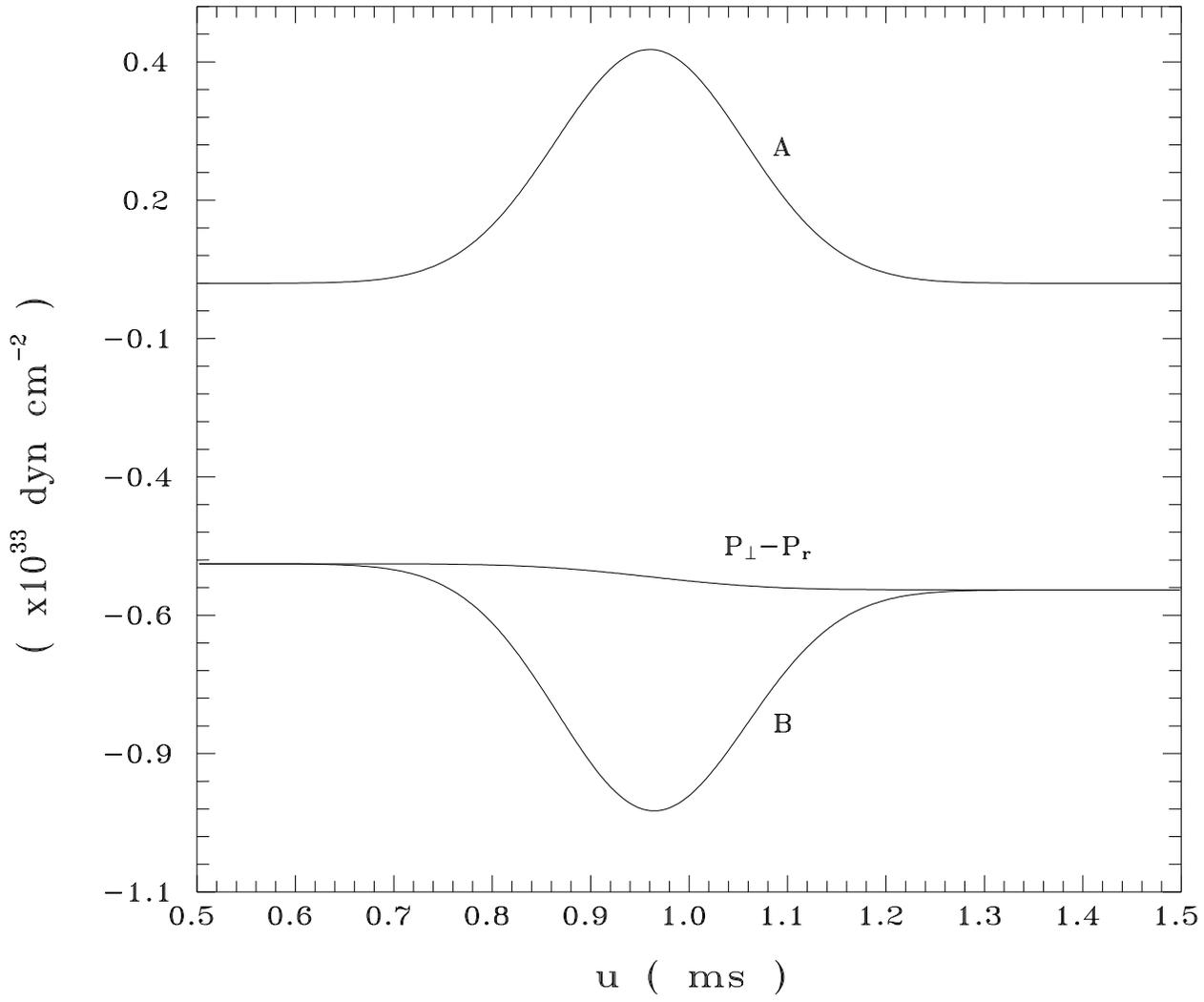
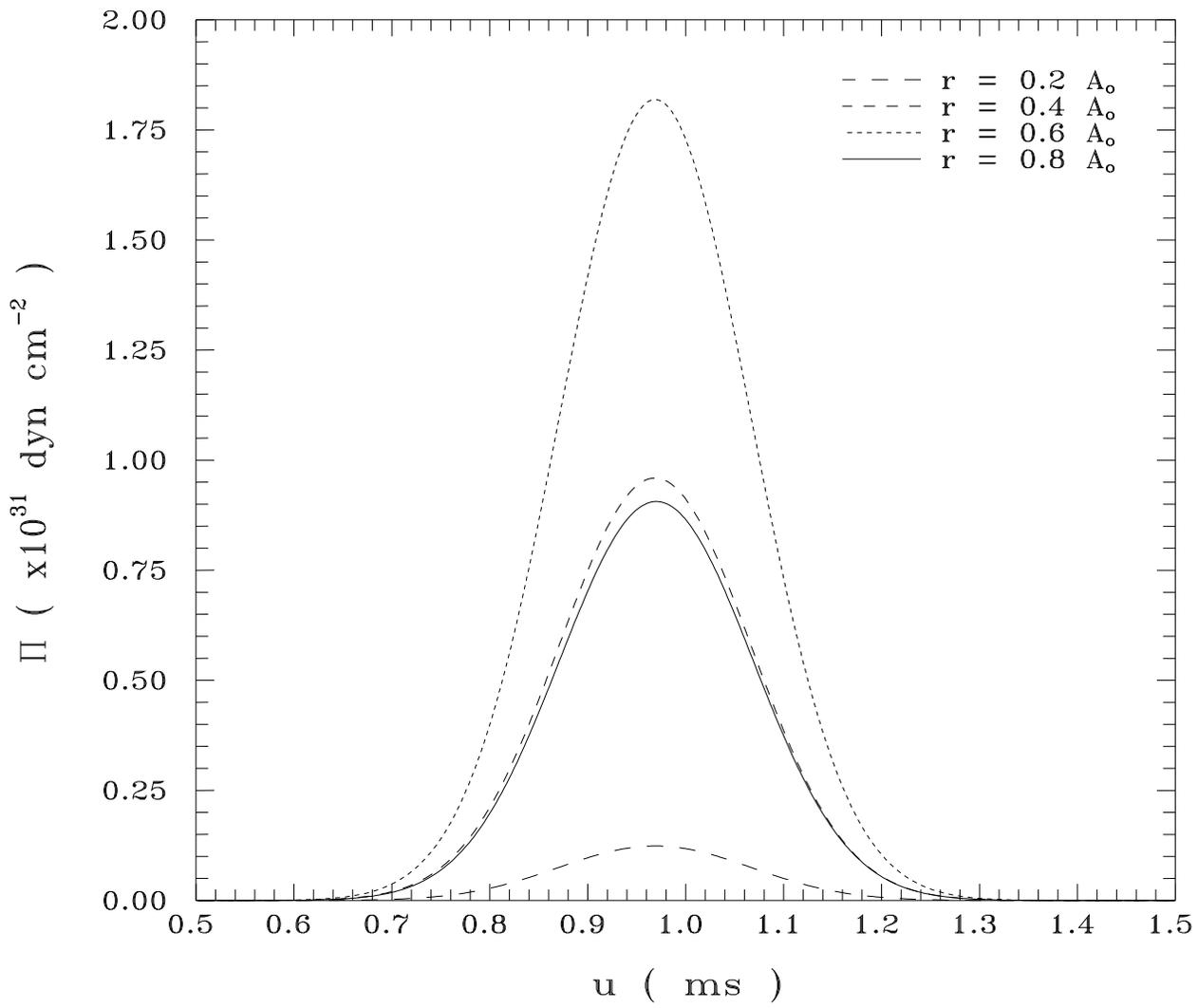
\end{document}